\documentclass[twocolumn,aps,prl,showpacs,amsmath,superscriptaddress,longbibliography,notitlepage]{revtex4-2}
\usepackage{amssymb}
\usepackage{mathrsfs}
\usepackage{graphicx}
\usepackage{float}
\usepackage[caption=false]{subfig}
\usepackage[pdftex,colorlinks=red]{hyperref}
\usepackage{verbatim}
\usepackage{txfonts}
\usepackage{epstopdf}
\usepackage[normalem]{ulem}
\usepackage{xcolor}
\usepackage{bm}
\usepackage[utf8]{inputenc}

\def\red{\textcolor{red}}

\begin{document}

\title{
{Anomalous hybridization of spectral winding topology in quantized steady-state responses}
}
\author{Hui-Qiang Liang}
\affiliation{Guangdong Provincial Key Laboratory of Quantum Metrology and Sensing $\&$ School of Physics and Astronomy, Sun Yat-Sen University (Zhuhai Campus), Zhuhai 519082, China}
\author{Sen Mu}\email{senmu@u.nus.edu}
\affiliation{Department of Physics, National University of Singapore, Singapore 117551, Republic of Singapore}
\author{Jiangbin Gong}\email{phygj@nus.edu.sg}
\affiliation{Department of Physics, National University of Singapore, Singapore 117551, Republic of Singapore}
\author{Linhu Li}\email{lilh56@mail.sysu.edu.cn}
\affiliation{Guangdong Provincial Key Laboratory of Quantum Metrology and Sensing $\&$ School of Physics and Astronomy, Sun Yat-Sen University (Zhuhai Campus), Zhuhai 519082, China}
\date{\today}

\begin{abstract}
Quantized response is one distinguishing feature of a topological system. In non-Hermitian systems, the spectral winding topology yields quantized steady-state response.  By considering two weakly coupled non-Hermitian chains,  we discover that the spectral winding topology of one chain can be probed by a steady-state response defined solely on the other chain, even when other important properties, e.g., {energetics} and entanglement entropy, indicate that eigen-solutions are effectively {not hybridized} between the two chains. This intriguing phenomenon, as carefully investigated in a large parameter space with a varying system size,  not only offers a new angle to understand interchain signal propagation in a non-Hermitian setting but also reveals unexpected physics of spectral winding topology vs quantized response. 
\end{abstract}

\maketitle

{\it Introduction.--} 
Rooted in their complex eigenenergies, the spectral winding topology of non-Hermitian systems constitutes a new aspect of topological physics with no Hermitian counterpart 
\cite{kawabata2019symmetry,shen2018topological,okuma2020topological,zhang2019correspondence}. 
Spectral winding is critical to predict the celebrated non-Hermitian skin effect (NHSE), as characterized by the localization of all eigenmodes at the system boundaries~\cite{alvarez2018non,yao2018edge,yokomizo2019non,song2019non,Lee2019hybrid,Lee2019anatomy,lee2020unraveling,li2020topological,yi2020nonH,helbig2020generalized,xiao2020non,ghatak2020observation}.
Nontrivial spectral winding has also been experimentally observed in an optical ring resonator with electro-optic modulators, where the complex spectrum is reproduced by mapping the lattice momentum to a frequency synthetic dimension \cite{wang2021generating}.
Further strengthening spectral winding as an important type of topology with unusual physical implications,
quantized steady-state responses were recently found in non-Hermitian chains,  with the height of the quantization plateaus determined by the winding numbers of a complex spectrum \cite{Li2021}.

Spectral winding is usually investigated under the periodic boundary conditions (PBCs), because it is simply prohibited under the open boundary conditions (OBCs) \cite{okuma2020topological,zhang2019correspondence}.  Near or at the limit of OBCs, other properties of non-Hermitian systems may change drastically upon tuning on an extremely weak boundary coupling \cite{budich2020sensor,mcdonald2020exponentially,li2021impurity,guo2021exact}, or a coupling between two chains with dissimilar skin localizations \cite{li2020critical,liu2020helical,mu2021nonhermitian}. Such hyper-sensitivity to weak perturbations are of continued interest, especially in connection with sensor designs \cite{jan2014enhancing,hodaei2017enhanced,chen2017exceptional,mcdonald2020exponentially,budich2020sensor,li2021impurity,guo2021exact,li2020critical,liu2020helical,mu2021nonhermitian}.
Under intermediate boundary conditions between PBCs and OBCs, some aspects of complex spectrum and eigenmode localization characteristics were also studied \cite{Lee2019anatomy,lee2020unraveling,li2021impurity,guo2021exact}.  This work investigates an unexplored problem, namely, how spectral winding features, as manifested in the quantized steady-state response, react to weak interchain coupling when a system is not under PBCs or OBCs.

Specifically, in examining quantized steady-state responses in connection with the spectral winding topology 
of two weakly coupled non-Hermitian chains, we discover a so-called {"anomalous hybridization" regime.
In this regime, on one hand the eigen-solutions are effectively {not hybridized} between the two chains in many aspects, e.g. vanishing entanglement between the two chains, quantized steady-state responses can be observed on either chain without complications due to the other; On the other hand, the spectral winding topology associated with one chain can be captured through a response defined solely on the other chain, indicating a nontrivial hybridization.} 
Such a counter-intuitive phenomenon is explained by considering the {propagation channels between the two chains under resonance condition.}
We then carefully investigate the transitions of the system between {dehybridization, anomalous hybridization and strong hybridization regimes}, through different quantized steady-state responses as well as an entanglement entropy analysis. 
Significantly, {the regime of anomalous hybridization}, which is of most interest here,  occurs in a relatively large parameter space, with the allowed interchain coupling much stronger than that to yield critical behaviors for systems initially under OBCs \cite{li2020critical}.  Our extensive computational results show that this regime further widens when increasing the system's size, indicating that {the physics of anomalous hybridization} uncovered here is even more typical in the thermodynamic limit.

\begin{figure}
\includegraphics[width=1\linewidth]{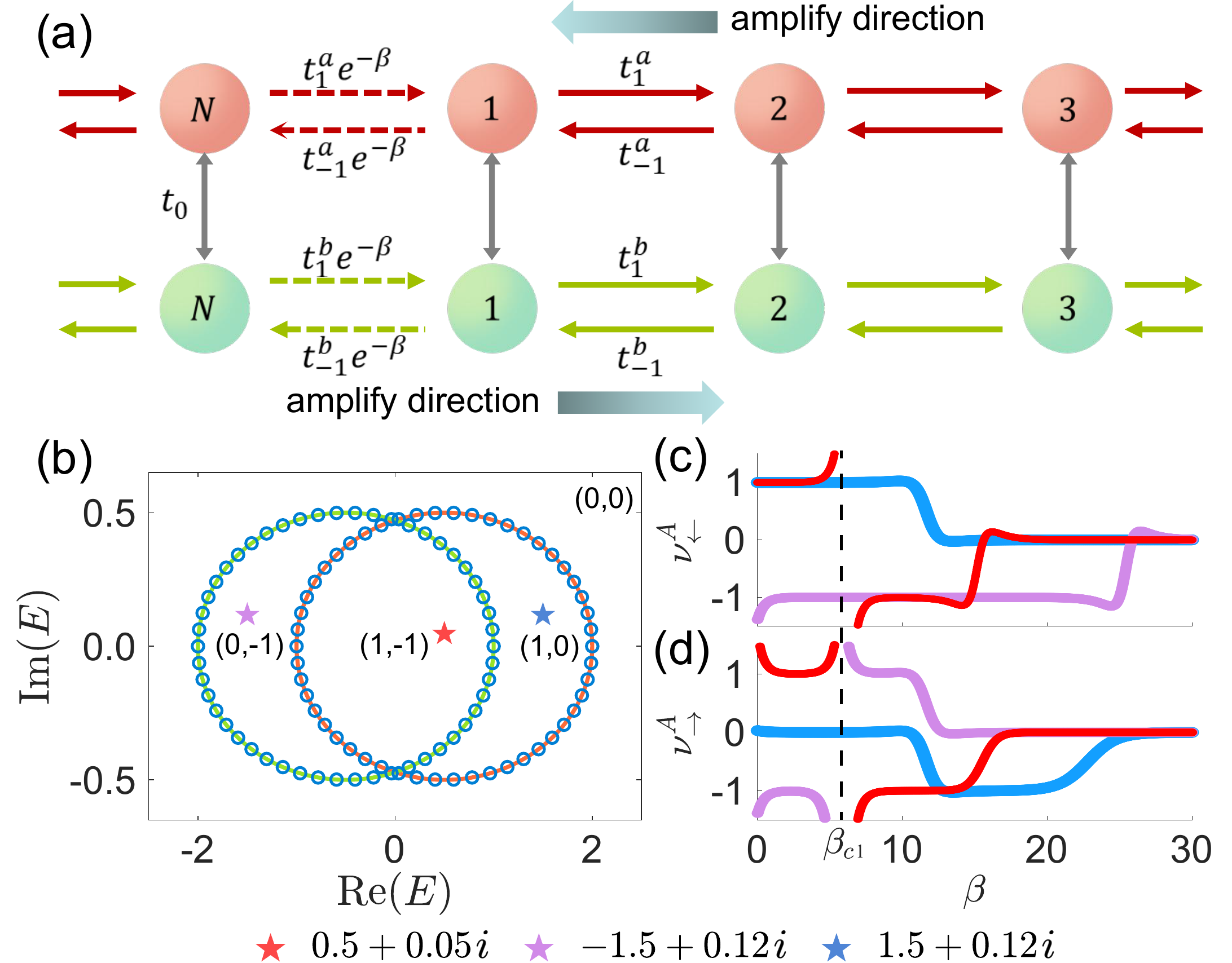}
\caption{(a) A sketch of the coupled-chain system.
(b) The PBC spectra of the system with $t_0=10^{-2.5}$ (blue dots) and $t_0=0$ (colored loops), which are almost on top to each other.
The PBC spectra of uncoupled chains in (a) with the same colors ($t_0=0$, solid loops), and that of the coupled system with $t_0=10^{-2.5}$.
The spectral winding numbers $\mathbf{w}(E_r)=(w_+(E_r),w_-(E_r))$ for $E_r$ in different regions are indicated in the figure.
(c) and (d) The quantized steady-state response quantities defined on chain $A$ (at $t_0=10^{-2.5}$), for the three chosen points labeled by stars with different colors in (b).
Other parameters are $N=50$, $t_1^A=_{-1}^B=1$, $t_{-1}^A=_{1}^B=0.5$, $V_A=-V_B=0.5$. 
}
\label{fig:fig1}
\end{figure}

{\it {Model.}--} 
We consider probing response on two weakly coupled Hatano-Nelson chains with different asymmetric nearest-neighbor (NN) hoppings, as shown in Fig.~\ref{fig:fig1}(a). The model system is described by the following tight-binding Hamiltonian:
\begin{equation}
\begin{aligned}
	\hat{H} =& \sum_{\alpha=A,B}\sum_{x=1}^{N-1}\left(t_1^\alpha{\hat{c}_{x,\alpha}}^\dag {\hat{c}_{x+1,\alpha}}+t_{-1}^\alpha{\hat{c}_{x+1,\alpha}}^\dag{\hat{c}_{x,\alpha}}\right)\\
	&+\sum_{x=1}^N\left(t_0{\hat{c}_{x,A}}^\dag \hat{c}_{x,B}+t_0{\hat{c}_{x,B}}^\dag \hat{c}_{x,A}+\sum_{\alpha=A,B}V^\alpha{\hat{c}_{x,\alpha}}^\dag \hat{c}_{x,\alpha}\right)\\
	&+e^{-\beta}\sum_{\alpha=A,B}\left(t_1^\alpha{\hat{c}_{N,\alpha}}^\dag \hat{c}_{1,\alpha} + t_{-1}^\alpha{\hat{c}_{1,\alpha}}^\dag \hat{c}_{N,\alpha}\right),
\end{aligned}\label{Hamiltonian}
\end{equation}
with $c^\dagger_{x,\alpha}$ the creation operator at the $x$-lattice site on chain-$\alpha$, $t^\alpha_{\pm1}$ the asymmetric hopping amplitudes and $V^\alpha$ the on-site potential in chain-$\alpha$, and $t_0$ the interchain coupling at each lattice site. 
When $t_0=0$, the non-Hermitian asymmetric hoppings of each chain leads to a directional amplification of a signal entering the system \cite{McDonald2018,Wanjura2020,xue2020non}, and in the following discussions we shall mainly focus on cases with the two chains having opposite amplification directions [as indicated in Fig.\ref{fig:fig1}(a)].
The boundary conditions in this model can be tuned from PBCs to OBCs by increasing $\beta$ from $0$ to infinity. Under PBCs, i.e, $\beta=0$, the Bloch Hamiltonian is given by 
\begin{eqnarray}
h(k)=\left(
\begin{array}{cc}
2t^A\cos (k-i\eta_A)+V^A & t_0\\
t_0 & 2t^B\cos  (k-i\eta_B)+V^B
\end{array}
\right),\label{eq_Hk}
\end{eqnarray}
with $k$ the quasi-momentum, {$\eta_{A/B}=\ln\sqrt{t^{A/B}_{1}/t^{A/B}_{-1}}$ the non-Hermitian inverse localization length, and $t^{A/B}=\sqrt{t^{A/B}_{1}t^{A/B}_{-1}}$.}
For either $\eta_{A/B}\neq0$, the Hamiltonian is non-Hermitian and point gapped with a topological winding number defined w.r.t. a reference energy $E_r$ in the complex energy plane,
\begin{eqnarray}
w(E_r)&=&w_+(E_r) +w_-(E_r),\label{eq:WN}
\end{eqnarray}
with
\begin{eqnarray}
w_{\pm}(E_r)&=&\int_{-\pi}^{\pi}\frac{dk}{2\pi i}\partial_{k}\ln\left[E_\pm(k)-E_r\right]
\label{eq:WN}
\end{eqnarray}
the single-band spectral winding numbers, $\pm$ the band index, and $E_\pm(k)$ the complex eigenenergies of the Hamiltonian.

Figure~\ref{fig:fig1} depicts the PBC spectrum of our system and its corresponding two-component winding numbers $$\mathbf{w}(E_r)=(w_+(E_r),w_-(E_r)).$$ {Assuming an ultra-weak interchain coupling $t_0\ll t^\alpha$, it is possible that the eigen-solutions of Eq.~\eqref{eq_Hk} are effectively not hybridized between the two chains.} Note that the two bands are seen to cross each other in the complex plane, yet we have $E_+(k)\neq E_-(k)$ for every lattice momentum $k$, thus the the notion of energy bands can still be well defined.
As seen in Fig. \ref{fig:fig1}(b), the two PBC bands with a small $t_0$ are virtually identical to the spectrum of the two {uncoupled} chains respectively. {We simply have $\mathbf{w}(E_r)=(w_A(E_r),w_B(E_r))$ at $t_0=0$.} 
As such, the single-band winding number $w_+(E_r)$ ($w_-(E_r)$) {is expected to reflect only the spectral winding topology of chain $A$ ($B$) under a weak interchain coupling.}
The complex energy plane is hence divided into four regimes, corresponding to different combinations of single-band winding numbers respectively.


{\it {Anomalous hybridization regime.}--} To physically manifest the spectral winding topology, we consider steady-state responses in a directional signal amplification process \cite{Li2021}. {The quantized coefficient is established from taking derivatives of a quantity involving Green's function w.r.t. the boundary tuning parameter $\beta$},
\begin{eqnarray}
\nu^{\alpha}_\leftarrow(E_r)=d\ln|G_{\alpha1,\alpha N}|/d\beta,
\nu^{\alpha}_\rightarrow(E_r)=d\ln|G_{\alpha N,\alpha1}|/d\beta,\label{eq:response}
\end{eqnarray}
with $\alpha$ representing either of the two chains,
and $G_{\alpha x,\alpha x'}$ an element of the Green's function $$G=1/(E_r-H)$$ associated with the $x$-th and $x'$-th site in chain-$\alpha$. $G_{\alpha x,\alpha x'}$ can describe a signal amplification between two sites on the same chain \cite{McDonald2018,Wanjura2020,xue2020non,Li2021}.
The reference energy can be expressed as $$E_r=\omega+i\gamma,$$ with $\omega$ the frequency of an input signal and $\gamma$ an extra uniform gain or loss adding to the system.
According to early results of quantized steady-state responses vs spectral winding topology {on a single chain} \cite{Li2021}, one expects to have $\nu^{\alpha}_\leftarrow(E_r)=w(E_r)$ for $w(E_r)=1$, or $\nu^{\alpha}_\rightarrow(E_r)=-w(E_r)$ for $w(E_r)=-1$. {These coefficients shall be non-positive otherwise, e.g. $\nu^{\alpha}_\rightarrow(E_r)\leq 0$ for $w(E_r)=1$, and always zero for large enough $\beta$ when the response itself becomes constant} \footnote{For $|\nu_\alpha|>1$, the single element shall be replaced by the determinant of its off-diagonal block when calculating the responses in Eq. \eqref{eq:response} \cite{Li2021}}.

We investigate the response on chain $A$ alone for our system. {Indeed, we obtain a plateau at $\nu_\leftarrow^A=1$ and non-positive $\nu_\rightarrow^A$ when $E_r$ falls in the region with $\mathbf{w}(E_r)=(1,0)$ (e.g. the blue star in Fig.~\ref{fig:fig1}(b)), as shown {by the blue lines} in Fig. \ref{fig:fig1}(c) and (d). This result is identical to the case when eigen-solutions are not hybridized. Next, for $E_r$ with $\mathbf{w}(E_r)=(0,-1)$ (e.g. the purple star in Fig.~\ref{fig:fig1}(b)), we see non-positive $\nu_\leftarrow^A$ in Fig.~\ref{fig:fig1}(c), and a plateau at $\nu_\rightarrow^A=1$ in Fig.~\ref{fig:fig1}(d) appear only after the boundary tuning parameter $\beta$ exceeds a certain value $\beta_{c1}$. This tells that the response $\nu_\rightarrow^A$ defined on chain $A$ {can only reflect the spectral winding related to chain $B$ when $\beta>\beta_{c1}$, since negative winding is contributed from chain $B$.} It should not be taken as a violation of the correspondence between winding topology and quantized steady-state response when $\beta<\beta_{c1}$, because here the response function is defined on chain $A$ only {(instead of involving both chains)}. {We thus do not expect it to predict the total winding number or $w_-(E_r)$ (from chain $B$) immediately away the PBCs, when the eigen-solutions are effectively not hybridized between the two chains yet.} In other words, $w_-(E_r)$ inherits the spectral winding topology of chain $B$ only and is not captured by the response $\nu^{A}_\rightarrow$ defined on chain $A$ for $E_r$ with $\mathbf{w}(E_r)=(0,-1)$ unless the eigen-solutions are hybridized when $\beta>\beta_{c1}$. It seems suggesting that the eigen-solutions in our system undergoes a sharp transition from not hybridized to hybridized at $\beta=\beta_{c1}$, and we will justify this conjecture from entanglement analysis between the two chains later.} 

Now let us consider $E_r$ falling in the central region with $\mathbf{w}(E_r) = (1,-1)$ (e.g. the red star in Fig.~\ref{fig:fig1}(b)). 
First, as seen in Fig.~\ref{fig:fig1}(c), $\nu_\leftarrow^A$ shows a plateau at $\nu_\leftarrow^A=1$ when $\beta<\beta_{c1}$, in agreement with the single-band winding number $w_+(E_r)=1$. For $\beta>\beta_{c1}$, $\nu_\leftarrow^A$ drops to non-positive values {due to hybridization of the eigen-solutions}, also in agreement with a total winding number $w(E_r)=0$ for the entire system. {Surprisingly, unlike for $E_r$ with $\mathbf{w}(E_r)=(0,-1)$, we find that the response $\nu^{A}_\rightarrow$ now gives rise to a plateau at $\nu^{A}_\rightarrow=1$ for $\beta<\beta_{c1}$, as shown by the red line in Fig.~\ref{fig:fig1}(d). This is unexpected since a positive $\nu^{A}_\rightarrow$ reflects a negative winding number which originates only from chain $B$ in our system. It unveils that the response $\nu^{A}_\rightarrow$ defined on chain $A$ can reflect the spectral winding topology of chain $B$ for $E_r$ with $\mathbf{w}(E_r)=(1,-1)$, though the eigen-solutions are effectively not hybridized between the two chains when $\beta<\beta_{c1}$ in our model.}
{Our results uncover another subtle aspect of coupled non-Hermitian chains: it is highly nontrivial to tell how the eigen-solutions cooperate in the steady-state response even when the interchain coupling is weak. Specifically, with several aspects in our system resembling to that of dehybridized eigen-solutions, a response defined on one chain can reveal the winding topology of the other for certain frequency of the input signal, clearly indicating a hybridization behaviour. 
To highlight this dual feature of dehybridization or hybridization in different measures, we dub it as ''anomalous hybridization" hereafter.} Similar behaviors are also seen in interchain response functions, associated with $G_{\alpha1,\bar{\alpha}N}$ and $G_{\alpha N,\bar{\alpha}1}$ where $\alpha\neq\bar{\alpha}$, as shown in the Supplemental Materials \cite{suppmat}.

\begin{figure}
\includegraphics[width=1\linewidth]{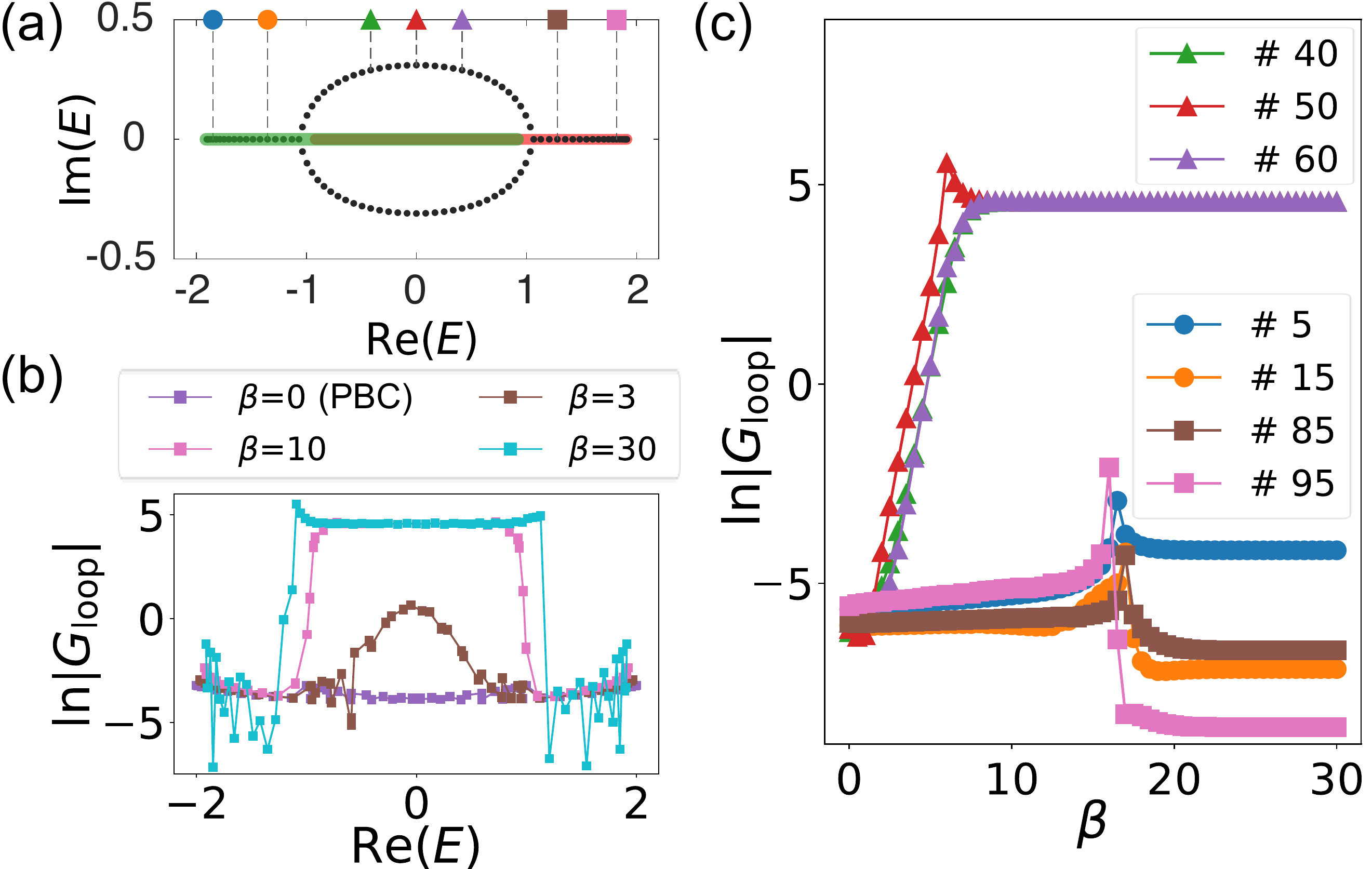}
\caption{(a) OBC spectra with $t_0=0$ (colored) and $t_0=10^{-2.5}$ (black). Red and green colors correspond to the spectra of two uncoupled chains respectively, overlapping in the central part of the spectrum. (b) $G_{\rm loop}=G_{A x,Bx'}G_{Bx',Ax}$ for reference energy $E_r$ chosen as eigenenergies of the system with a small imaginary energy detuning in calculating the Green's function.
Parameters adopted are $N=50,x=20,x'=30$. (c) $G_{\rm loop}$ for several different eigenenergies labeled in (a) with the same symbols and colors. Legend shows the orders of the eigenenergies sorted in their real parts.
Parameters are chosen to be the same as for Fig.~\ref{fig:fig1}.
}
\label{fig:fig2}
\end{figure}

{\it Propagation channels between two coupled chains.--} 
The seemingly self-contradictory phenomenon above may be qualitatively understood as the following:  though the eigen-solutions are effectively not hybridized, propagation channels between the two chains are always present, allowing the information to possibly "propagate" from one chain to the other. To verify this picture, we first consider the system under OBCs with $\beta\rightarrow \infty$, and employ the Green's function to examine {the interchain propagation in the bulk of the system, instead of the end-to-end response on one chain alone.}
Note that when the two chains are uncoupled, spectra of the two chains under OBCs are real and partially overlapping due to their on-site potentials $V^{A,B}$ [see Fig. \ref{fig:fig2}(a)]. Upon turning on the interchain coupling $t_0$, a {signal} at one chain with frequency $\omega$ might propagate to the other chain. {Such propagation is favored when both chains have eigenmodes sharing almost the same eigenenergy close to $\omega$, satisfying the resonance condition. Therefore, an element from the Green's function related to the interchain response  shall be much larger for the input frequency $\omega$ falling in the central part of the spectrum than that for the frequency at the tails.}
This argument is confirmed by our numerical results for a loop response defined by
$$G_{\rm loop}=G_{A x,Bx'}G_{B x',Ax},$$
representing {the product of two amplification ratios, one for a signal traveling from site $x$ in chain $A$ to site $x'$ in chain $B$ and the other vice versa, as shown in Fig. \ref{fig:fig2}(b).} {More importantly, once a signal propagates to the other chain, it can be directionally amplified there, thus carrying over the winding topology of that chain before propagating back.} Besides, these interchain propagating channels effectively form a propagation loop, and the signal travelling along the loop can be amplified repeatedly. This amplification mechanism can be interpreted as instabilities as well, hence the emergence of complex eigenenergies in the central part of the OBC spectrum, as shown in 
Fig.~\ref{fig:fig2}(a). 

{We now discuss why the boundary tuning parameter $\beta$ makes a difference in our observations for $G_{\rm loop}$, as seen in Fig.~\ref{fig:fig2}(b) and (c).}  First note that $G_{\rm loop}$ is related to signal amplification between two distanced lattice sites, which shall vanish under the PBCs \cite{Wanjura2020,xue2020non}. 
Intuitively, boundary couplings $t^{A,B}_{\pm 1}e^{-\beta}$ connecting each chain head to tail also provide intrachain propagation channels, which {may be enhanced by the non-reciprocal pumping and overwhelm the interchain ones under weak interchain coupling, i.e. $t_0<t^{A,B}_{\pm 1}e^{-\beta}$.}
Indeed, we see $G_{\rm loop}$ is almost vanishing when the system is under the PBCs ($\beta=0$), as shown in Fig.~\ref{fig:fig2}(b). 
On the other hand, when the system is tuned away from the PBCs (i.e. increasing $\beta$ from zero), $G_{\rm loop}$ becomes larger 
only for $\omega$ falling in the central part of the spectrum, {where interchain propagations are favored due to resonances.} In Fig.~\ref{fig:fig2}(c), we present $G_{\rm loop}$ for several different eigenenergies, and it is clearly seen that $G_{\rm loop}$ increases rapidly with $\beta$, for eigenenergies in the central part of the spectrum.
That is, for the parameters we consider, the propagation channels {start to play a role} when the system is slightly tuned away from the PBCs.
The existence of such propagation channels allows the spectral topology of one chain, which is essentially a property under PBCs, to be unveiled from the steady-state response defined on the other chain {for certain energy window, even when the eigen-solutions are effectively not hybridized between the two chains.}

\begin{figure}
\includegraphics[width=1\linewidth]{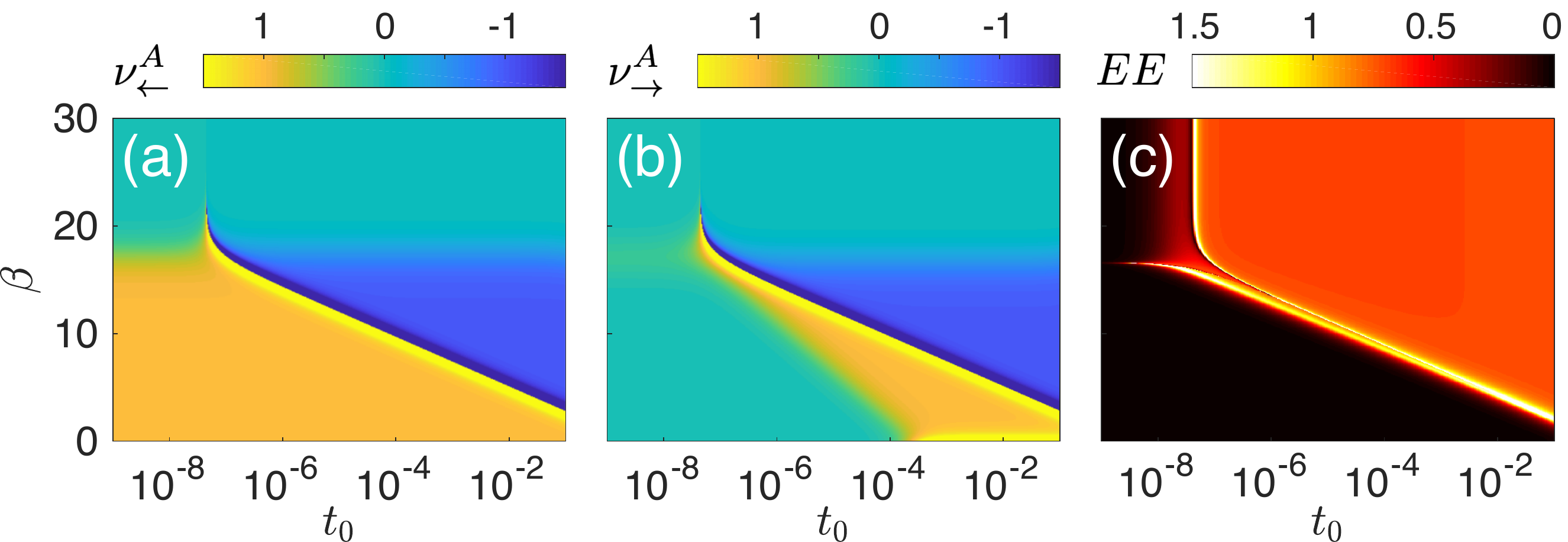}
\caption{Quantized steady-state responses defined on the $A$ chain (a) $\nu_{\leftarrow}^A$ and (b) $\nu_{\rightarrow}^A$, for a reference energy at $E_r=0$. {The yellow area in (a) has a steady-state response with $\nu_{\leftarrow}^A=1$, reflecting only the single-band winding number $w_+(0)$.
Eigen-solutions are effectively dehybridized between the two chains, which is further verified by the vanishing EE in (c).
In (b), a subarea (yellow) of the dehybridization regime gives a steady-state response with $\nu_{\rightarrow}^A=1$,
suggesting an anomalous hybridization behavior with resonant interchain propagation channels.}
(c) Entanglement entropy (EE) for the $N$th eigenmode sorted in its real energy.
Other parameters are $t_1^A=_{-1}^B=1$, $t_{-1}^A=_{1}^B=0.5$, $V_A=-V_B=0.5$, and the system's size is chosen to be $N=50$.
}
\label{fig:fig3}
\end{figure}

{\it Topological response and entanglement entropy.--} 
It remains to study in more depth the properties of the interchain propagation channels in the competition between interchain coupling $t_0$ and the boundary tuning parameter $\beta$.
Figures~\ref{fig:fig3}(a) and (b) present the steady-state response defined in Eq.~\eqref{eq:response} on chain $A$ alone for the reference energy $E_r=0$, i.e. the center of the spectrum with $\mathbf{w}(E_r)=(1,-1)$. {One sees $\nu^A_{\leftarrow}=1$ (yellow regime) for a wide range of $t_0$ when $\beta$ is below certain $\beta_{c1}$ in Fig.~\ref{fig:fig3}(a). {This suggests that the eigen-solutions are effectively not hybridized between the two chains.} The other response $\nu^A_{\rightarrow}$, shown in Fig.~\ref{fig:fig3}(b), is also quantized at $\nu^A_{\rightarrow}=1$ (yellow regime), reflecting that the spectral winding of chain $B$ can be probed from chain $A$ in a subregime of the yellow regime in Fig.~\ref{fig:fig3}(a). We term the yellow regime in Fig.~\ref{fig:fig3}(b) as anomalous hybridization, which covers a rather large parameter space. Anomalous hybridization signifies that interchain propagation channels are in favor, though eigen-solutions are effectively not hybridized between the two chains. Besides, note that with a larger $\beta$, the system approaches the OBC limit with trivial spectral winding, and hence always gives trivial steady-state response. With this insight and previous results \cite{koch2020bulk,kunst2018biorthogonal,li2021impurity}, we infer that the steady-state response considered here can distinguish between dehybridization and anomalous hybridization regimes when $\beta<\beta_{\rm OBC}\approx N\ln \sqrt{t^A_1/t^A_{-1}}$.}
 
{To further justify the dehybridization of the eigen-solutions in the anomalous hybridization regime and verify our conjecture on the emergence of $\beta_{c1}$, we investigate the interchain entanglement entropy (EE) and compare the results with the quantized responses obtained.} 
Specifically,
we compute the biorthogonal EE for an entanglement cut chosen between the two chains, defined as 
\cite{chang2020entanglement,li2020critical,li2021non}
\begin{eqnarray}
S_n=-\sum_m\eta_{n,m}\ln \eta_{n,m}+(1-\eta_{n,m})\ln(1-\eta_{n,m}),
\end{eqnarray}
where $\eta_{n,m}$ is the $m$-th eigenvalue of the correlator matrix $C_n$ for the $n$-th eigenmode,
\begin{eqnarray}
(C_n)_{xy}=\langle\Psi^L_n|\hat{c}^\dagger_{x}\hat{c}_{y} |\Psi^R_n\rangle
\end{eqnarray}
with $x,y$ from only chain $A$.
In Fig.~\ref{fig:fig3}(c) we show the EE for the $N$th eigenmode (sorted in its real energy), which corresponds to the most pronounced $G_{\rm loop}$ in Fig. \ref{fig:fig2}(b) and (c).
{The regime with $S_N\approx0$ coincides well with the dehybridization regime, whereas the regime with $S_N\approx \ln 2$ corresponds to the hybridization regime found in Fig.~\ref{fig:fig3}(a). Notably, these two said regimes are seen to have sharp boundaries. That is, in the anomalous hybridization regime identified as the yellow area in Fig.~\ref{fig:fig3}(b), entanglement between the two chains is vanishing. This confirms again that the eigen-solutions are effectively not hybridized.} 

\begin{figure}
\includegraphics[width = 1\linewidth]{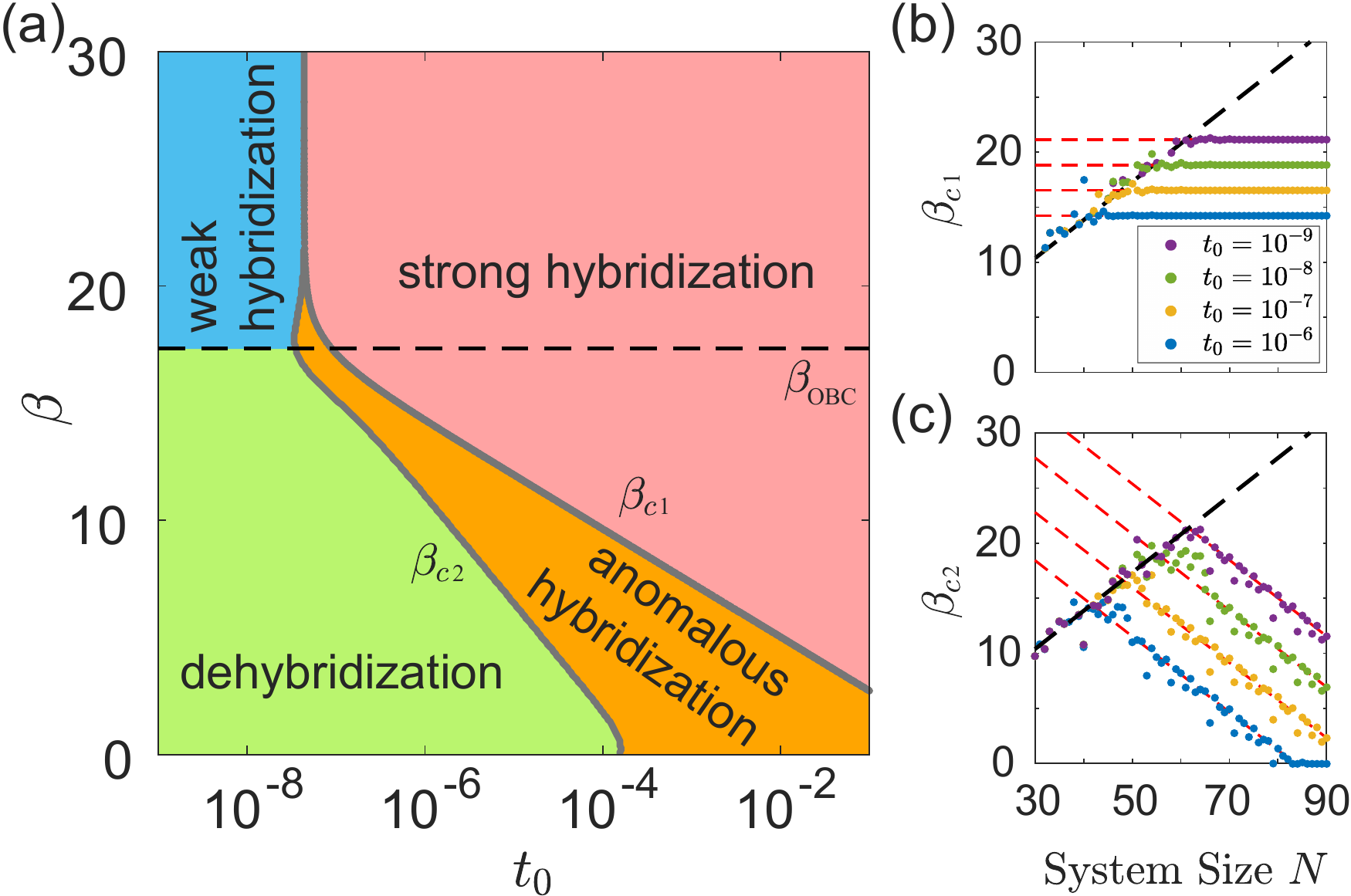}
\caption{(a) A phase diagram of the system read out from the quantized response quantities and EE in Fig. \ref{fig:fig3}. {Colors indicate different phases as labeled in the figure.}
(b) and (c) show how the phase boundaries of $\beta_{c1}$ and $\beta_{c2}$ varies with the system's size $N$, with the black dash lines indicating $\beta_{\rm OBC}\approx N\ln\sqrt{t_1^A/t_{-1}^A}$, where the system approaches the OBC limit.
The data points are read out from the jumps of the quantized response quantities $\nu_{\rightarrow}^A$. 
Red dash lines are the numerical fitting of the data points.
The three critical values for a fixed $N$ roughly cross each other at the same point, as further shown in the Supplemental Materials \cite{suppmat}. Parameters are $t_1^A = t_{-1}^B = 1$, $t_{-1}^A = t_1^B = 0.5$, and $V_A = -V_B = 0.5$.
Numerically we obtain 
$\beta_{c1}\approx \ln t_0$,
which is also independent from the reference energy $E_r$ \cite{suppmat}.
}\label{fig:fig4}
\end{figure}

{\it Phase diagram.--} 
{We finally present a phase diagram of our system based on steady-state response and entanglement analysis in Fig.~\ref{fig:fig4}(a). The dehybridization regime is represented by the green area, where single-chain spectral winding topology manifests itself as a quantized steady-state response on that chain alone. The anomalous hybridization regime is marked by the orange area, where eigen-solutions are effectively not hybridized (vanishing entanglement between the two chain), while importantly interchain propagation channels allow the spectral winding topology of one chain to be detected from the response defined on the other chain alone. The pink area represents a strong hybridization regime, where the interchain EE is saturated, and the steady-state response in this regime reflects the total spectral winding of the system for a reference energy when $\beta<\beta_{\rm OBC}$.} Last, the blue area depicts a weak hybridization regime, where 
{EE decreases with $t_0$ as shown in Fig.~\ref{fig:fig3}(c).} 

{Our phase diagram highlights several transitions when increasing $\beta$ from $0$,  
e.g. with $t_0= 10^{-5}$ for $E_r$ with $\mathbf{w}(E_r)=(1,-1)$, 
(i) a transition between anomalous hybridization regime and strong hybridization at $\beta_{c1}$ 
(ii) a transition between dehybridization regime and anomalous hybridization 
at $\beta_{c2}$.
Another critical value $\beta_{\rm OBC}$ is known to be proportional to the system's size $N$ \cite{koch2020bulk,kunst2018biorthogonal,li2021impurity}, which indicates the system behaves more like under OBC with spectral winding vanishing.} Interestingly,
$\beta_{c1}$ and $\beta_{c2}$ are found to exhibit markedly different scaling with the system's size $N$, as shown in Fig.~\ref{fig:fig4}(b) and (c).
That is, 
$\beta_{c1}$ remains a constant as $N$ varies, 
whereas $\beta_{c2}$ decreases when increasing $N$ (see the Supplemental Materials \cite{suppmat} for more details).
In summary, it is now evident that as system's size increases, the anomalous hybridization regime (orange regime) bordered by the lines $\beta_{c2}$ and $\beta_{c1}$ widens. {Hence the anomalous hybridization regime is approachable in a larger parameter space when the system is taken to the thermodynamic limit. Indeed, it cannot be stressed enough that non-Hermitian systems can be extremely sensitive to couplings between boundaries or between different subsystems.}
The Supplemental Materials \cite{suppmat} contains more detailed analysis of $\beta_{c2}$ and $\beta_{c1}$ by considering different specific values of the reference energy $E_r$. 

{\it Conclusion.--} 
Topological physics in non-Hermitian systems continues to bring us surprises. In investigating the probe of spectral winding topology through quantized steady-state responses to changes in boundary coupling, we unveil a counter-intuitive phenomenon in a {non-Hermitian coupled chain setting.  Even when the observed energetics and the entanglement entropy clearly indicate that the eigen-solutions are effectively not hybridized,} the spectral winding topology of one chain can be probed by a response defined on the other chain alone for certain frequency of the input signal. {Termed as anomalous hybridization, this was explained by the dominance of the resonant interchain propagation channels over the intrachain ones.} 
While enhancing the notion that the topology of spectral winding on the complex energy plane can be physically manifested as quantized steady-state responses, our results have revealed unexpected possibilities in connecting spectral winding topology with physical measurements. 

{\it Acknowledgements.--}
H. L. and S. M. contribute equally to this work. L. L. acknowledges funding support by the National Natural Science Foundation of China (12104519) and the Guangdong Basic and Applied Basic Research Foundation (2020A1515110773), J. G. acknowledges support from Singapore National Research Foundation Grant No. NRF- NRFI2017-04 (WBS No. R-144-000-378-281).

\clearpage

\onecolumngrid
\begin{center}
\textbf{\large Supplementary Materials}\end{center}
\setcounter{equation}{0}
\setcounter{figure}{0}
\renewcommand{\theequation}{S\arabic{equation}}
\renewcommand{\thefigure}{S\arabic{figure}}
\renewcommand{\cite}[1]{\citep{#1}}

\section{1. Further results of the quantized steady-state response}
In Fig. 1 in the main text we have shown the behavior of the quantized quantities $\nu^{A}_\leftarrow(E_r)$ and $\nu^{A}_\rightarrow(E_r)$ in the so-called anomalous hybridization regime, 
indicating the domination of interchain propagation channels even when the eigen-solutions of the system are seemingly dehybridized. 
By definition, these two quantities correspond to the elements $G_{A1,An}$ and $G_{An,A1}$ of the Green's function, reflecting the signal amplification between the two ends of chain $A$.
For completeness, in this supplemental material we shall display more results in different scenarios and compare them with the results in the main text.

\begin{figure}[H]
\centering
\includegraphics[width = 0.9\textwidth]{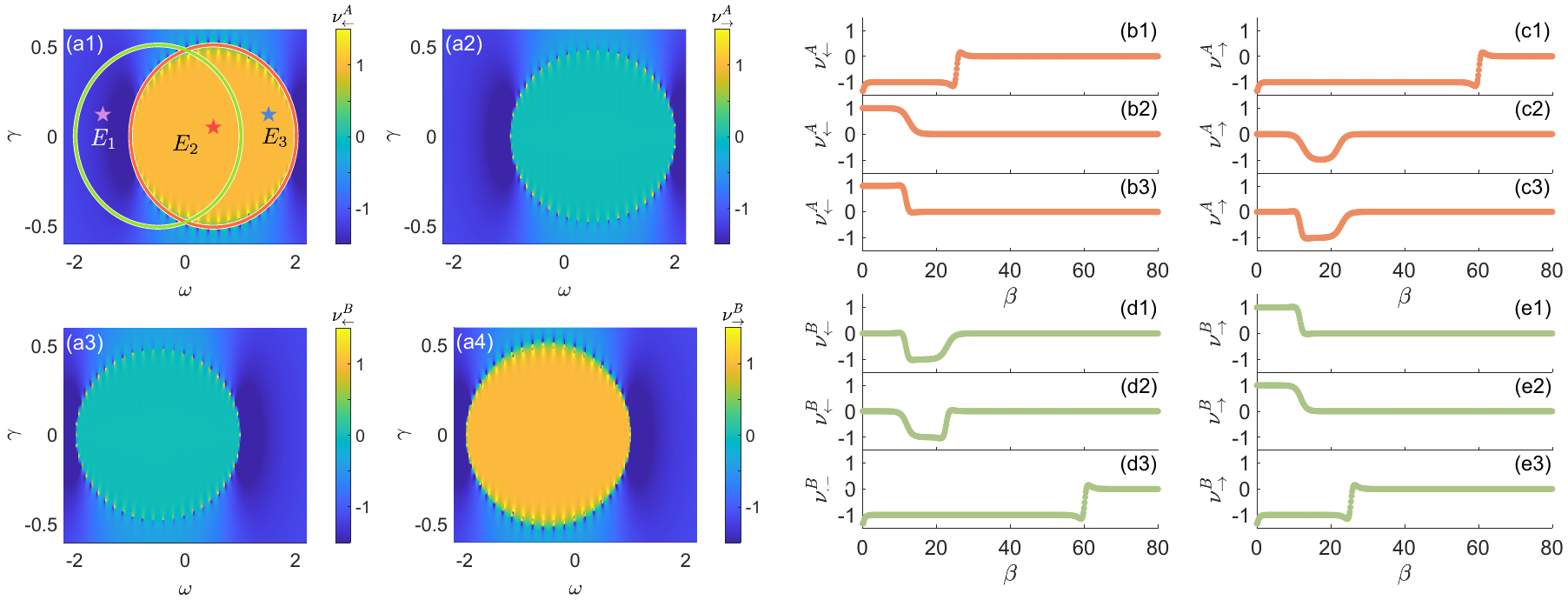}
\caption{
The quantized steady-state response for two decoupled chains. (a) The four response quantities defined for chain $A$ and $B$ respectively at the PBCs with $\beta=0$. In (a1) we also demonstrate the spectra of the two chains (orange and green loops for chain $A$ and $B$), and the three reference energies for obtaining the results in (b) to (e), $E_1=-1.5+0.12i$, $E_2=0.5+0.05i$, and $E_3=1.5+0.12i$.
(b)-(e) the four response quantities as functions of $\beta$. 
In each panel, the three sub-panels labeled with numbers show the results obtained for the reference energies with the same subscript numbers in (a1).
Other parameters are $N=50$, $t_1^A=_{-1}^B=1$, $t_{-1}^A=_{1}^B=0.5$, $V_A=-V_B=0.5$, the same as those in Fig. 1 in the main text.  
}
\label{supp_fig1}
\end{figure}

\subsection{Steady-state response for decoupled chains}
We first consider the quantized steady-state responses, defined in Eqs. 5 in the main text, for two decoupled chains at $t_0=0$.
In Fig. \ref{supp_fig1}(a) we illustrate the four response quantities at $\beta=0$ (under the PBCs), which shall correspond to the spectral winding topology under the periodic boundary conditions (PBCs).
It is seen that $\nu^A_{\leftarrow}(E_r)=1$ in the area enclosed by the PBC spectrum of chain $A$ [orange loop in Fig. \ref{supp_fig1}(a1)], corresponding to its spectral winding number, $w_A(E_r)=1$ for $E_r$ within this area.
Meanwhile, the other response quantity $\nu^A_{\rightarrow}(E_r)$ in (a2) does not reflect this positive spectral winding number \cite{Li2021}.
When $\beta$ increases, 
the response quantity $\nu^A_{\leftarrow}$ gives a plateau at $\nu^A_{\leftarrow}=1$ before it jumps to zero at a critical value of $\beta$,
only for reference energies $E_r$ enclosed by the PBC spectrum of chain $A$, 
as shown in Fig. \ref{supp_fig1}(b).
Otherwise $\nu^A_{\leftarrow}$ remains non-positive for arbitrary $\beta$ (except at some critical values), and so does $\nu^A_{\rightarrow}$ shown in Fig. \ref{supp_fig1}(c).
Since the two chains are now decoupled, $\nu^A_{\leftarrow}$ and $\nu^A_{\rightarrow}$ do not tell the spectral winding topology of chain $B$, whose spectrum is given by the green loop in Fig. \ref{supp_fig1}(a1).

Similarly, the response quantities defined on chain $B$, $\nu^B_{\leftarrow}$ and $\nu^B_{\rightarrow}$, indicates only the spectral winding topology of chain $B$, as shown in Fig. \ref{supp_fig1}(a3), (a4), (d), and (e). Note that compared to chain $A$, the results of the these two quantities exchange since the spectral winding of chain $B$ is $w_B(E_r)=-1$ for the parameters we choose, which is reflected by $\nu^B_{\rightarrow}$ \cite{Li2021}.

\subsection{Steady-state response between the two chains}
\begin{figure}
\centering
\includegraphics[width = 1\textwidth]{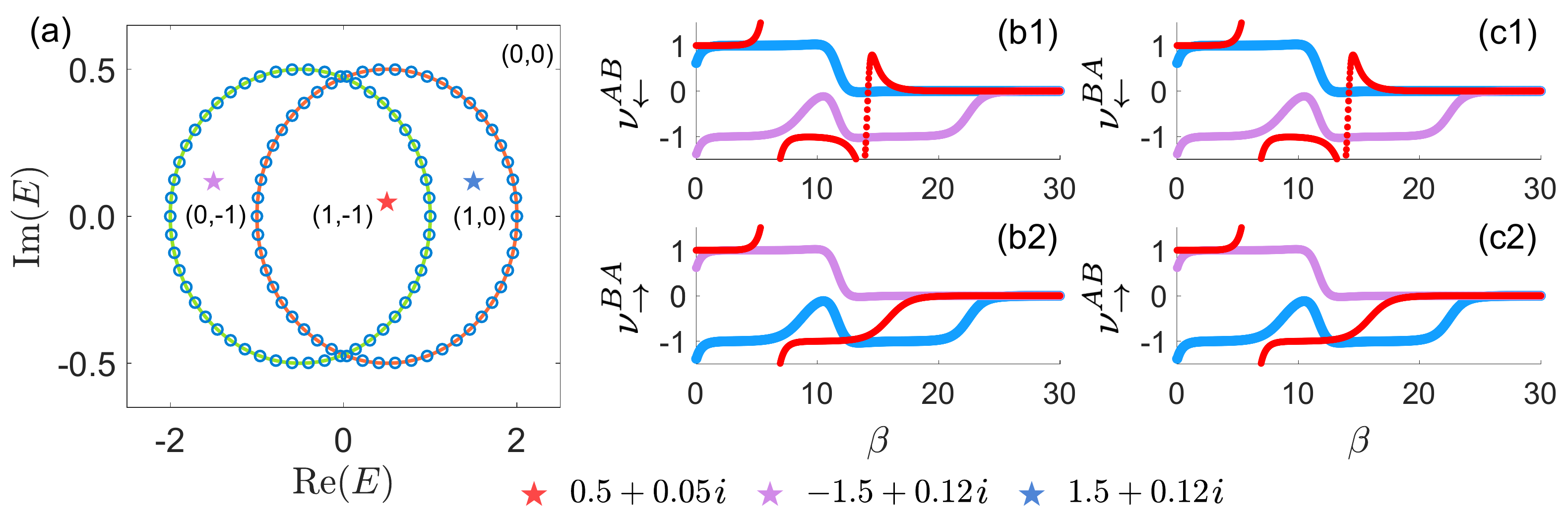}
\caption{(a) the same PBC spectra as that in Fig. 1 in the main text, with $t_0=10^{-2.5}$ (blue dots) and $t_0=0$ (colored loops).
The PBC spectra of the two decoupled chains in (a) with the same colors ($t_0=0$, solid loops), and that of the coupled system with $t_0=10^{-2.5}$.
The spectral winding numbers $\mathbf{w}(E_r)=(w_+(E_r),w_-(E_r))$ for $E_r$ in different regions are indicated in the figure.
(b) and (c) the quantized steady-state response quantities between the two chains, defined as in Eqs. \ref{eq:supp_responseAB}, for the reference energies $E_r$ at the stars with the same colors in (a).
Other parameters are $N=50$, $t_1^A=_{-1}^B=1$, $t_{-1}^A=_{1}^B=0.5$, $V_A=-V_B=0.5$. 
}
\label{supp_fig2}
\end{figure}
{Following the discussion in the main text, the discovered anomalous hybridization can be understood as a result of the domination of resonant propagation channels between the two effectively decoupled chains. Therefore we also expect to see similar behavior of the response associated with the elements $G_{\alpha 1,\bar{\alpha} n}$ and $G_{\alpha  n,\bar{\alpha}  1}$ with $\alpha\neq\bar{\alpha}$, which describes the end-to-end signal amplification between different chains. 
To this end, we define response quantities as
\begin{eqnarray}
\nu_{\leftarrow}^{\alpha\bar{\alpha}}(E_r)=d\ln|G_{\alpha1,\bar{\alpha} N}|/d\beta,~~~~
\nu_{\rightarrow}^{\alpha\bar{\alpha}}(E_r)=d\ln|G_{\alpha N,\bar{\alpha} 1}|/d\beta,\label{eq:supp_responseAB}
\end{eqnarray}
and display our numerical results in Fig. \ref{supp_fig2}.
It is seen that for a reference energy with $\mathbf{w}=(1,-1)$ (red star in the figure), these interchain response quantities exhibit the same plateaux at $1$ in the anomalous hybridization regime with small $\beta$. They jump to a non-positive value ($-1$) afterward, where the eigen-solutions of the two chains are strongly hybridized, and the responses reflect the total winding number $w(E_r)=1-1=0$.
For the other two regimes where $E_r$ enclosed  by the spectrum of one band [e.g. the blue and purple stars in Fig. \ref{supp_fig2}(a)], the interchain response quantities are seen to reflect corresponding winding numbers. That is, $\nu_{\leftarrow}^{\alpha\bar{\alpha}}$ ($\nu_{\rightarrow}^{\alpha\bar{\alpha}}$) gives a quantized plateaux at $1$ only when the winding number is $1$ ($-1$), and jumps to non-positive value when the system is strongly hybridized.}

{Note that here the interchain response never reflects the dehybridized behavior discussed in the main text, where the intrachain response of one chain does not reflect the winding number of the other chain (see the discussion about the purple star for Fig. 1 in the main text). This is because for the interchain response defined here, a signal has to travel through the interchain propagation channels to get to the other end. In other words, in this treatment we have post-selected only the signal propagating between the two chains, and hence the response quantities alway reflect the winding numbers of the system. }

\subsection{Steady-state response in other parameter regimes}
\begin{figure}
\centering
\includegraphics[width = 0.9\textwidth]{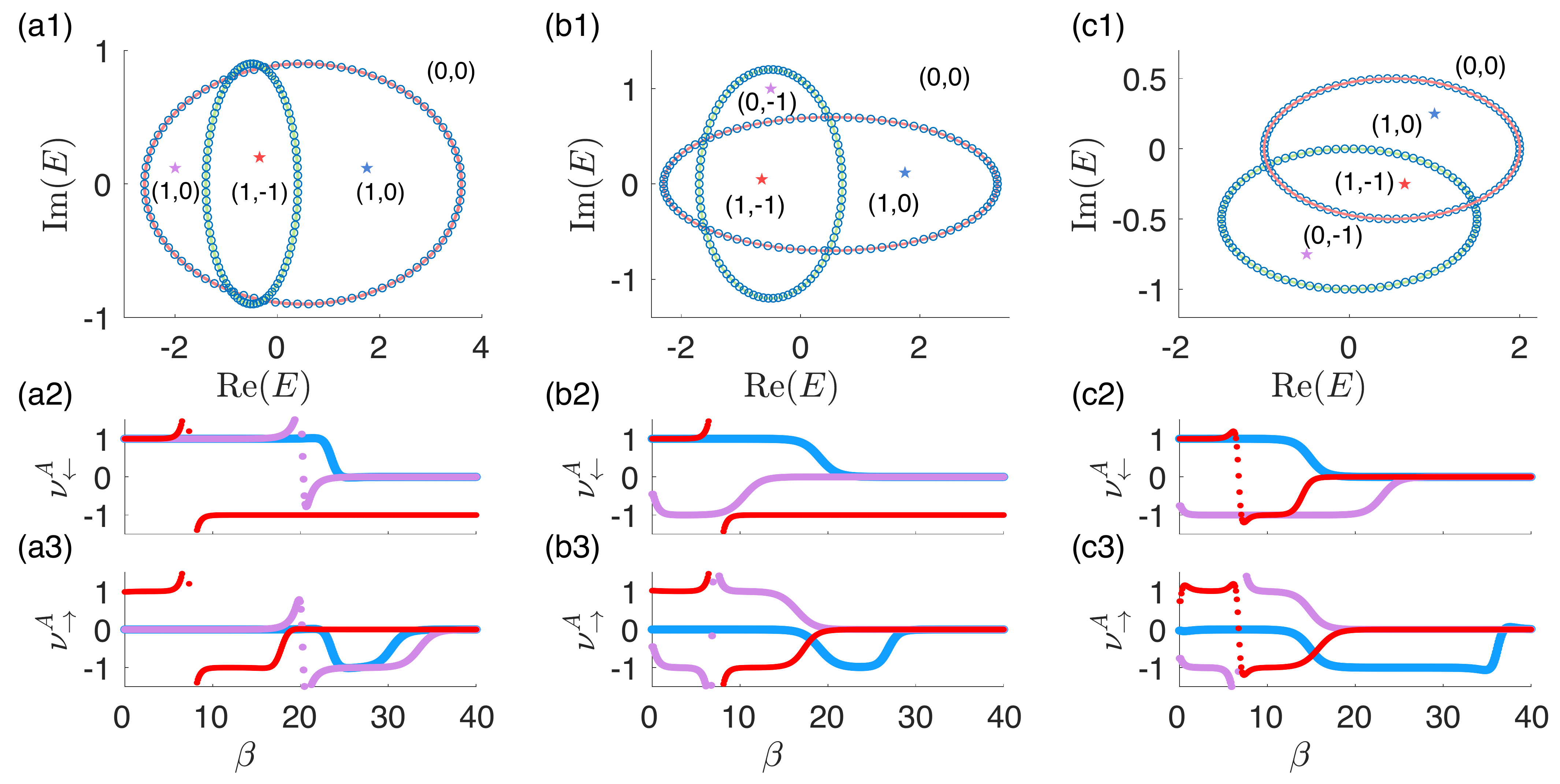}
\caption{ PBC spectra and the quantized steady-state responses for more general cases.
Parameters are $t_0=10^{-3}$, and
(a) $t_1^A=2$, $t_{-1}^A=1.1$, $t_1^B=0$, $t_{-1}^B=0.9$, $V_A=-V_B=0.5$;
(b) $t_1^A=1.75$, $t_{-1}^A=1.05$, $t_1^B=0$, $t_{-1}^B=1.2$, $V_A=-V_B=0.5$;
(c) $t_1^A=1$, $t_{-1}^A=0.5$, $t_1^B=0.5$, $t_{-1}^B=1$, $V_A=-iV_B=0.5$.
}
\label{supp_fig3}
\end{figure}

In the main text we have consider a case with $t_1^A=t_{-1}^B$ and $t_{-1}^A=t_1^B$, i.e. the two chains have the same non-Hermitian pumping strength, but toward opposite directions. In Fig.\ref{supp_fig3} we display several examples with parameters tuned away from this symmetric regime, and the quantized steady-state responses can also be clearly seen in these cases. 
In Fig.\ref{supp_fig3} (a), the response quantities of purple and blue stars behave similarly, as they are both enclosed only by the same spectral loop and have $\mathbf{w}(E_r)=(1,0)$.
In Fig.\ref{supp_fig3} (b) and (c), the purple and blue stars correspond to different spectral winding numbers, which are also reflected by the response quantities.
In either case, when $E_r$ falls in the area enclosed by both loops, the response defined on chain $A$ reflects both single-band spectral winding numbers $w_\pm(E_r)=\pm1$ when $\beta$ is small, in consistent with our results presented in the main text.


\section{2. Further results about the phase boundaries of $\beta_{c1}$ and $\beta_{c2}$}
\begin{figure}
\includegraphics[width = 0.9\textwidth]{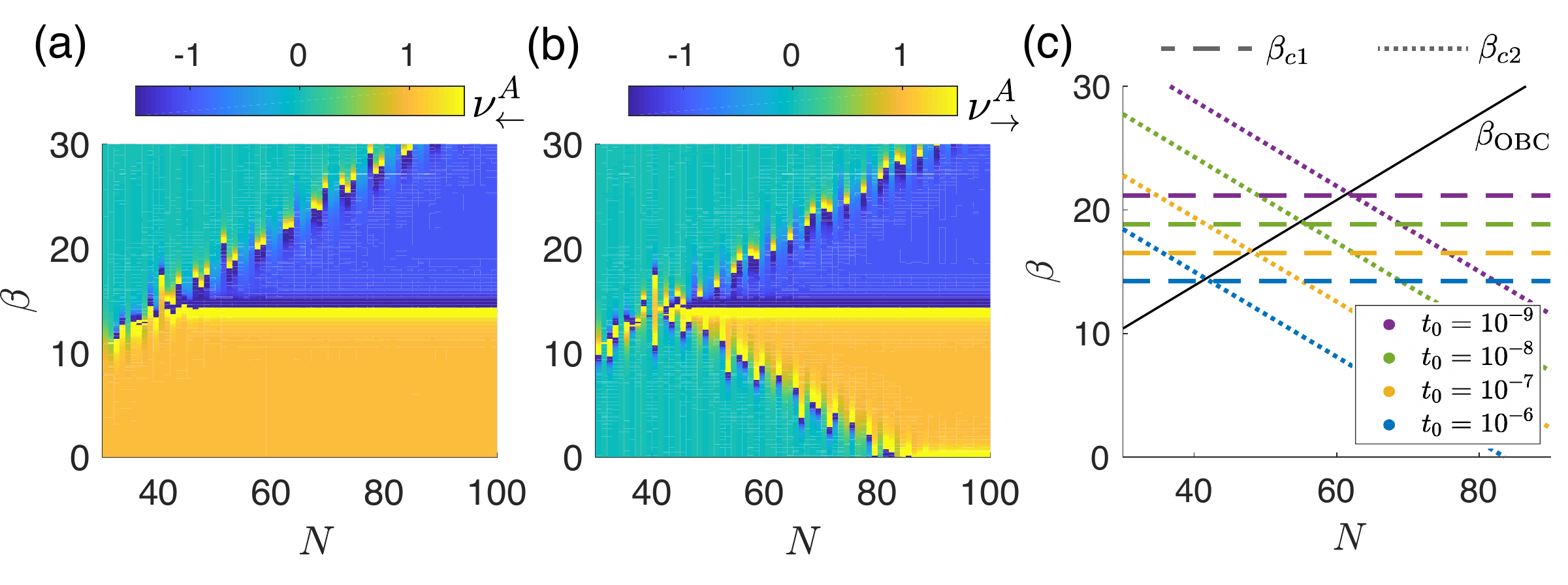}
\caption{Size-dependence (-independence) of $\beta_{c2}$ ($\beta_{c1}$). 
The two quantized response quantities $\nu^A_{\leftarrow}$ and $\nu^A_{\rightarrow}$ with $t_0=10^{-0.6}$ are displayed in (a) and (b), for different values of $\beta$ and the system's size $N$. The reference energy is chose as $E_r=0$, the same as the red star in Fig. 3 in the main text.
Combining these results, there are four phases with different $(\nu^A_{\leftarrow},\nu^A_{\rightarrow})$. separated by the three boundaries of $\beta_{c2}$, $\beta_{c1}$, and $\beta_{\rm OBC}$.
(c) The three boundaries versus $N$ for different $t_0$. The blue dash and doted lines are read out from (b), and the others are read out from similar results with different interchain coupling strengths.
Other parameters are $t_1^A=_{-1}^B=1$, $t_{-1}^A=_{1}^B=0.5$, $V_A=-V_B=0.5$,
}
\label{supp_fig4}
\end{figure}

As shown in the main text, the anomalous hybridization with interchain propagation channels occurs in the regime with $\beta_{c2}<\beta<\beta_{c1}$. In Fig. \ref{supp_fig4}(a) and (b), we illustrate the two quantized response quantities $\nu^A_{\leftarrow}$ and $\nu^A_{\rightarrow}$, with the same parameters as that in Fig. 3 in the main text. The single-band winding numbers $\mathbf{w}=(1,-1)$ are reflected by the two response quantities in the yellow areas in the two panels respectively, and the one in Fig. \ref{supp_fig4}(b) gives the anomalous hybridization regime. Three phase boundaries are clearly seen in Fig. \ref{supp_fig4}(b), corresponding to $\beta_{c2}$, $\beta_{c1}$, and $\beta_{\rm OBC}$ from lower to top respectively, as discussed in the main text.
Fitting of these numerical data points for these boundaries are shown in Fig. \ref{supp_fig4}(c), displaying the size-dependence (-independence) nature of $\beta_{c2}$ ($\beta_{c1}$. Notably, for different strengths of the interchain couplings, the three boundaries are always seen to cross each other at the same point.

\begin{figure}
\includegraphics[width = 1\textwidth]{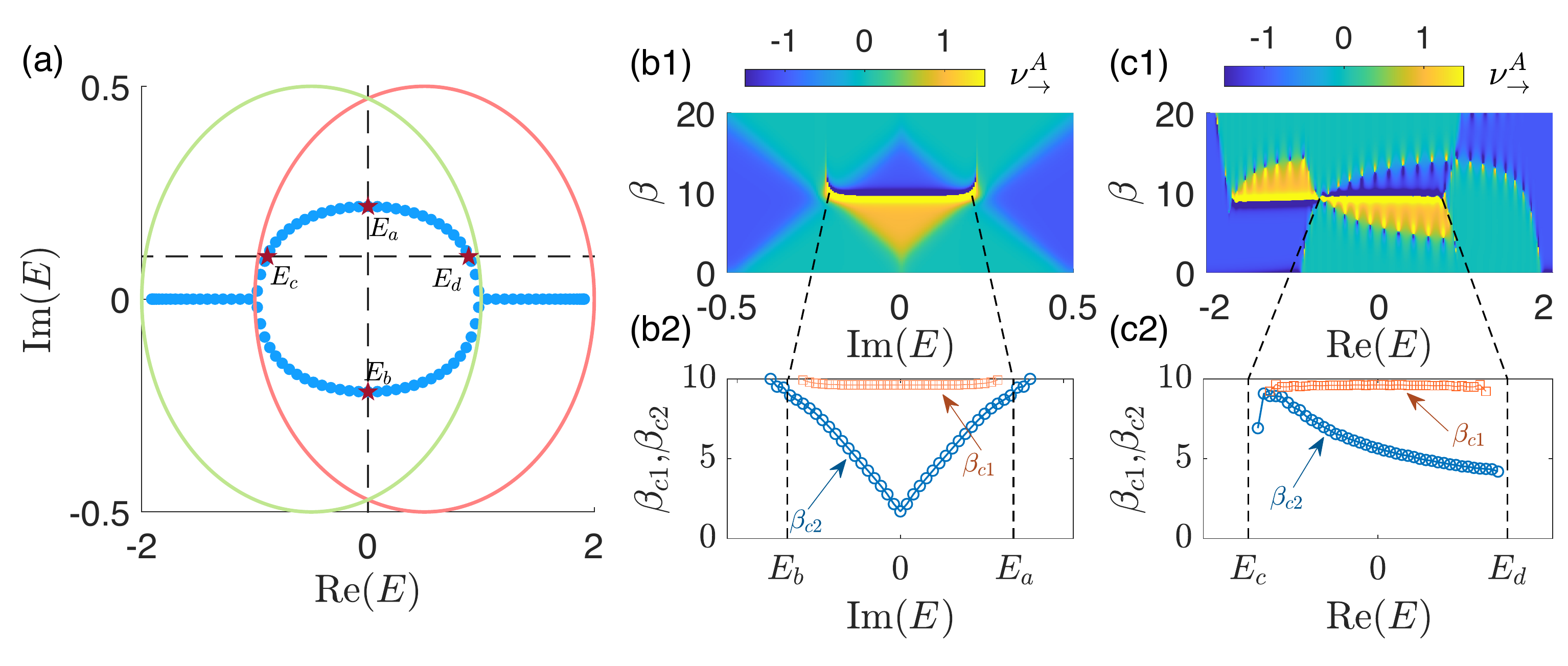}
\caption{ $E_r$-independence (-dependence) of $\beta_{c1}$ ($\beta_{c2}$). (a) PBC (green and pink loops) and OBC spectra of the system.
(b) The response quantity $\nu_{\rightarrow}^A$ for $E_r$ along the vertical dash line in (a). (b2) shows the critical values $\beta_{c1}$ and $\beta_{c2}$ read out from (b1).
(c) The response quantity $\nu_{\rightarrow}^A$ for $E_r$ along the horizontal dash line in (a). (c2) shows the critical values $\beta_{c1}$ and $\beta_{c2}$ read out from (c1). The left yellow area in (c1) corresponds to where the coupling is strong enough to hybridize the two chains, and the winding numbers are $\mathbf{w}=(0,-1)$, so that $\nu_{\rightarrow}^A$ reflects the absolute value of the total winding number $w=0-1=-1$.
}
\label{supp_fig5}
\end{figure}

In Fig. \ref{supp_fig5} we display the numerical results of $\beta_{c1}$ and $\beta_{c2}$ for different reference energy $E_r$.
Specifically, we have chosen $E_r$ along the two dash lines in Fig. \ref{supp_fig5}(a), and display the corresponding response quantity $\nu^A_{\rightarrow}$ in Fig. \ref{supp_fig5}(b1) and (c1). The anomalous hybridization regime is given by the central yellow area in each figure, where the reference energy $E_r$ falls in the central ring of the OBC spectrum in Fig. \ref{supp_fig5}(a).
The two critical values $\beta_{c1}$ and $\beta_{c2}$ are thus read out in these areas, and $\beta_{c1}$ is seen to be independent from $E_r$, as shown in Fig. \ref{supp_fig5}(b2) and (c2). 

To give an explanation of why the anomalous hybridization occurs only for $E_r$ enclosed by the central ring of the OBC spectrum, 
we note that the it is rooted in the interchain propagation channels, 
which forms a closed path for a signal to grow or decay exponentially when traveling along it. The growing/decaying rate of the signal correspond to the imaginary part of the OBC eigenenergies.
However, for $E_r$ outside the central ring of the OBC spectrum, its imaginary part has a larger amplitude than the OBC eigeneneriges (with the same real part as that of $E_r$),
effectively describes a system with a local gain/loss rate stronger than that of the closed path formed by the interchain propagation channels.
In other words, the system is now dominated by this local gain/loss, {overwhelming any signal traveling through the interchain propagation channels, and hence the system behaves as not hybridized.}

\begin{thebibliography}{37}%
\makeatletter
\providecommand \@ifxundefined [1]{%
 \@ifx{#1\undefined}
}%
\providecommand \@ifnum [1]{%
 \ifnum #1\expandafter \@firstoftwo
 \else \expandafter \@secondoftwo
 \fi
}%
\providecommand \@ifx [1]{%
 \ifx #1\expandafter \@firstoftwo
 \else \expandafter \@secondoftwo
 \fi
}%
\providecommand \natexlab [1]{#1}%
\providecommand \enquote  [1]{``#1''}%
\providecommand \bibnamefont  [1]{#1}%
\providecommand \bibfnamefont [1]{#1}%
\providecommand \citenamefont [1]{#1}%
\providecommand \href@noop [0]{\@secondoftwo}%
\providecommand \href [0]{\begingroup \@sanitize@url \@href}%
\providecommand \@href[1]{\@@startlink{#1}\@@href}%
\providecommand \@@href[1]{\endgroup#1\@@endlink}%
\providecommand \@sanitize@url [0]{\catcode `\\12\catcode `\$12\catcode
  `\&12\catcode `\#12\catcode `\^12\catcode `\_12\catcode `\%12\relax}%
\providecommand \@@startlink[1]{}%
\providecommand \@@endlink[0]{}%
\providecommand \url  [0]{\begingroup\@sanitize@url \@url }%
\providecommand \@url [1]{\endgroup\@href {#1}{\urlprefix }}%
\providecommand \urlprefix  [0]{URL }%
\providecommand \Eprint [0]{\href }%
\providecommand \doibase [0]{https://doi.org/}%
\providecommand \selectlanguage [0]{\@gobble}%
\providecommand \bibinfo  [0]{\@secondoftwo}%
\providecommand \bibfield  [0]{\@secondoftwo}%
\providecommand \translation [1]{[#1]}%
\providecommand \BibitemOpen [0]{}%
\providecommand \bibitemStop [0]{}%
\providecommand \bibitemNoStop [0]{.\EOS\space}%
\providecommand \EOS [0]{\spacefactor3000\relax}%
\providecommand \BibitemShut  [1]{\csname bibitem#1\endcsname}%
\let\auto@bib@innerbib\@empty
\bibitem [{\citenamefont {Kawabata}\ \emph {et~al.}(2019)\citenamefont
  {Kawabata}, \citenamefont {Shiozaki}, \citenamefont {Ueda},\ and\
  \citenamefont {Sato}}]{kawabata2019symmetry}%
  \BibitemOpen
  \bibfield  {author} {\bibinfo {author} {\bibfnamefont {K.}~\bibnamefont
  {Kawabata}}, \bibinfo {author} {\bibfnamefont {K.}~\bibnamefont {Shiozaki}},
  \bibinfo {author} {\bibfnamefont {M.}~\bibnamefont {Ueda}},\ and\ \bibinfo
  {author} {\bibfnamefont {M.}~\bibnamefont {Sato}},\ }\bibfield  {title}
  {\bibinfo {title} {Symmetry and topology in non-hermitian physics},\
  }\href@noop {} {\bibfield  {journal} {\bibinfo  {journal} {Physical Review
  X}\ }\textbf {\bibinfo {volume} {9}},\ \bibinfo {pages} {041015} (\bibinfo
  {year} {2019})}\BibitemShut {NoStop}%
\bibitem [{\citenamefont {Shen}\ \emph {et~al.}(2018)\citenamefont {Shen},
  \citenamefont {Zhen},\ and\ \citenamefont {Fu}}]{shen2018topological}%
  \BibitemOpen
  \bibfield  {author} {\bibinfo {author} {\bibfnamefont {H.}~\bibnamefont
  {Shen}}, \bibinfo {author} {\bibfnamefont {B.}~\bibnamefont {Zhen}},\ and\
  \bibinfo {author} {\bibfnamefont {L.}~\bibnamefont {Fu}},\ }\bibfield
  {title} {\bibinfo {title} {Topological band theory for non-hermitian
  hamiltonians},\ }\href@noop {} {\bibfield  {journal} {\bibinfo  {journal}
  {Phys. Rev. Lett.}\ }\textbf {\bibinfo {volume} {120}},\ \bibinfo {pages}
  {146402} (\bibinfo {year} {2018})}\BibitemShut {NoStop}%
\bibitem [{\citenamefont {Okuma}\ \emph {et~al.}(2020)\citenamefont {Okuma},
  \citenamefont {Kawabata}, \citenamefont {Shiozaki},\ and\ \citenamefont
  {Sato}}]{okuma2020topological}%
  \BibitemOpen
  \bibfield  {author} {\bibinfo {author} {\bibfnamefont {N.}~\bibnamefont
  {Okuma}}, \bibinfo {author} {\bibfnamefont {K.}~\bibnamefont {Kawabata}},
  \bibinfo {author} {\bibfnamefont {K.}~\bibnamefont {Shiozaki}},\ and\
  \bibinfo {author} {\bibfnamefont {M.}~\bibnamefont {Sato}},\ }\bibfield
  {title} {\bibinfo {title} {Topological origin of non-hermitian skin
  effects},\ }\href {https://doi.org/10.1103/PhysRevLett.124.086801} {\bibfield
   {journal} {\bibinfo  {journal} {Phys. Rev. Lett.}\ }\textbf {\bibinfo
  {volume} {124}},\ \bibinfo {pages} {086801} (\bibinfo {year}
  {2020})}\BibitemShut {NoStop}%
\bibitem [{\citenamefont {Zhang}\ \emph {et~al.}(2020)\citenamefont {Zhang},
  \citenamefont {Yang},\ and\ \citenamefont {Fang}}]{zhang2019correspondence}%
  \BibitemOpen
  \bibfield  {author} {\bibinfo {author} {\bibfnamefont {K.}~\bibnamefont
  {Zhang}}, \bibinfo {author} {\bibfnamefont {Z.}~\bibnamefont {Yang}},\ and\
  \bibinfo {author} {\bibfnamefont {C.}~\bibnamefont {Fang}},\ }\bibfield
  {title} {\bibinfo {title} {Correspondence between winding numbers and skin
  modes in non-hermitian systems},\ }\href
  {https://doi.org/10.1103/PhysRevLett.125.126402} {\bibfield  {journal}
  {\bibinfo  {journal} {Phys. Rev. Lett.}\ }\textbf {\bibinfo {volume} {125}},\
  \bibinfo {pages} {126402} (\bibinfo {year} {2020})}\BibitemShut {NoStop}%
\bibitem [{\citenamefont {Alvarez}\ \emph {et~al.}(2018)\citenamefont
  {Alvarez}, \citenamefont {Vargas},\ and\ \citenamefont
  {Torres}}]{alvarez2018non}%
  \BibitemOpen
  \bibfield  {author} {\bibinfo {author} {\bibfnamefont {V.~M.}\ \bibnamefont
  {Alvarez}}, \bibinfo {author} {\bibfnamefont {J.~B.}\ \bibnamefont
  {Vargas}},\ and\ \bibinfo {author} {\bibfnamefont {L.~F.}\ \bibnamefont
  {Torres}},\ }\bibfield  {title} {\bibinfo {title} {Non-hermitian robust edge
  states in one dimension: Anomalous localization and eigenspace condensation
  at exceptional points},\ }\href@noop {} {\bibfield  {journal} {\bibinfo
  {journal} {Phys. Rev. B}\ }\textbf {\bibinfo {volume} {97}},\ \bibinfo
  {pages} {121401} (\bibinfo {year} {2018})}\BibitemShut {NoStop}%
\bibitem [{\citenamefont {Yao}\ and\ \citenamefont {Wang}(2018)}]{yao2018edge}%
  \BibitemOpen
  \bibfield  {author} {\bibinfo {author} {\bibfnamefont {S.}~\bibnamefont
  {Yao}}\ and\ \bibinfo {author} {\bibfnamefont {Z.}~\bibnamefont {Wang}},\
  }\bibfield  {title} {\bibinfo {title} {Edge states and topological invariants
  of non-hermitian systems},\ }\href
  {https://doi.org/10.1103/PhysRevLett.121.086803} {\bibfield  {journal}
  {\bibinfo  {journal} {Phys. Rev. Lett.}\ }\textbf {\bibinfo {volume} {121}},\
  \bibinfo {pages} {086803} (\bibinfo {year} {2018})}\BibitemShut {NoStop}%
\bibitem [{\citenamefont {Yokomizo}\ and\ \citenamefont
  {Murakami}(2019)}]{yokomizo2019non}%
  \BibitemOpen
  \bibfield  {author} {\bibinfo {author} {\bibfnamefont {K.}~\bibnamefont
  {Yokomizo}}\ and\ \bibinfo {author} {\bibfnamefont {S.}~\bibnamefont
  {Murakami}},\ }\bibfield  {title} {\bibinfo {title} {Non-bloch band theory of
  non-hermitian systems},\ }\href@noop {} {\bibfield  {journal} {\bibinfo
  {journal} {Physical review letters}\ }\textbf {\bibinfo {volume} {123}},\
  \bibinfo {pages} {066404} (\bibinfo {year} {2019})}\BibitemShut {NoStop}%
\bibitem [{\citenamefont {Song}\ \emph {et~al.}(2019)\citenamefont {Song},
  \citenamefont {Yao},\ and\ \citenamefont {Wang}}]{song2019non}%
  \BibitemOpen
  \bibfield  {author} {\bibinfo {author} {\bibfnamefont {F.}~\bibnamefont
  {Song}}, \bibinfo {author} {\bibfnamefont {S.}~\bibnamefont {Yao}},\ and\
  \bibinfo {author} {\bibfnamefont {Z.}~\bibnamefont {Wang}},\ }\bibfield
  {title} {\bibinfo {title} {Non-hermitian skin effect and chiral damping in
  open quantum systems},\ }\href@noop {} {\bibfield  {journal} {\bibinfo
  {journal} {Physical review letters}\ }\textbf {\bibinfo {volume} {123}},\
  \bibinfo {pages} {170401} (\bibinfo {year} {2019})}\BibitemShut {NoStop}%
\bibitem [{\citenamefont {Lee}\ \emph {et~al.}(2019)\citenamefont {Lee},
  \citenamefont {Li},\ and\ \citenamefont {Gong}}]{Lee2019hybrid}%
  \BibitemOpen
  \bibfield  {author} {\bibinfo {author} {\bibfnamefont {C.~H.}\ \bibnamefont
  {Lee}}, \bibinfo {author} {\bibfnamefont {L.}~\bibnamefont {Li}},\ and\
  \bibinfo {author} {\bibfnamefont {J.}~\bibnamefont {Gong}},\ }\bibfield
  {title} {\bibinfo {title} {Hybrid higher-order skin-topological modes in
  nonreciprocal systems},\ }\href
  {https://doi.org/10.1103/PhysRevLett.123.016805} {\bibfield  {journal}
  {\bibinfo  {journal} {Phys. Rev. Lett.}\ }\textbf {\bibinfo {volume} {123}},\
  \bibinfo {pages} {016805} (\bibinfo {year} {2019})}\BibitemShut {NoStop}%
\bibitem [{\citenamefont {Lee}\ and\ \citenamefont
  {Thomale}(2019)}]{Lee2019anatomy}%
  \BibitemOpen
  \bibfield  {author} {\bibinfo {author} {\bibfnamefont {C.~H.}\ \bibnamefont
  {Lee}}\ and\ \bibinfo {author} {\bibfnamefont {R.}~\bibnamefont {Thomale}},\
  }\bibfield  {title} {\bibinfo {title} {Anatomy of skin modes and topology in
  non-hermitian systems},\ }\href {https://doi.org/10.1103/PhysRevB.99.201103}
  {\bibfield  {journal} {\bibinfo  {journal} {Phys. Rev. B}\ }\textbf {\bibinfo
  {volume} {99}},\ \bibinfo {pages} {201103} (\bibinfo {year}
  {2019})}\BibitemShut {NoStop}%
\bibitem [{\citenamefont {Lee}\ \emph {et~al.}(2020)\citenamefont {Lee},
  \citenamefont {Li}, \citenamefont {Thomale},\ and\ \citenamefont
  {Gong}}]{lee2020unraveling}%
  \BibitemOpen
  \bibfield  {author} {\bibinfo {author} {\bibfnamefont {C.~H.}\ \bibnamefont
  {Lee}}, \bibinfo {author} {\bibfnamefont {L.}~\bibnamefont {Li}}, \bibinfo
  {author} {\bibfnamefont {R.}~\bibnamefont {Thomale}},\ and\ \bibinfo {author}
  {\bibfnamefont {J.}~\bibnamefont {Gong}},\ }\bibfield  {title} {\bibinfo
  {title} {Unraveling non-hermitian pumping: Emergent spectral singularities
  and anomalous responses},\ }\href
  {https://doi.org/10.1103/PhysRevB.102.085151} {\bibfield  {journal} {\bibinfo
   {journal} {Phys. Rev. B}\ }\textbf {\bibinfo {volume} {102}},\ \bibinfo
  {pages} {085151} (\bibinfo {year} {2020})}\BibitemShut {NoStop}%
\bibitem [{\citenamefont {Li}\ \emph {et~al.}(2020{\natexlab{a}})\citenamefont
  {Li}, \citenamefont {Lee},\ and\ \citenamefont {Gong}}]{li2020topological}%
  \BibitemOpen
  \bibfield  {author} {\bibinfo {author} {\bibfnamefont {L.}~\bibnamefont
  {Li}}, \bibinfo {author} {\bibfnamefont {C.~H.}\ \bibnamefont {Lee}},\ and\
  \bibinfo {author} {\bibfnamefont {J.}~\bibnamefont {Gong}},\ }\bibfield
  {title} {\bibinfo {title} {Topological switch for non-hermitian skin effect
  in cold-atom systems with loss},\ }\href@noop {} {\bibfield  {journal}
  {\bibinfo  {journal} {Physical Review Letters}\ }\textbf {\bibinfo {volume}
  {124}},\ \bibinfo {pages} {250402} (\bibinfo {year}
  {2020}{\natexlab{a}})}\BibitemShut {NoStop}%
\bibitem [{\citenamefont {Yi}\ and\ \citenamefont {Yang}(2020)}]{yi2020nonH}%
  \BibitemOpen
  \bibfield  {author} {\bibinfo {author} {\bibfnamefont {Y.}~\bibnamefont
  {Yi}}\ and\ \bibinfo {author} {\bibfnamefont {Z.}~\bibnamefont {Yang}},\
  }\bibfield  {title} {\bibinfo {title} {Non-hermitian skin modes induced by
  on-site dissipations and chiral tunneling effect},\ }\href
  {https://doi.org/10.1103/PhysRevLett.125.186802} {\bibfield  {journal}
  {\bibinfo  {journal} {Phys. Rev. Lett.}\ }\textbf {\bibinfo {volume} {125}},\
  \bibinfo {pages} {186802} (\bibinfo {year} {2020})}\BibitemShut {NoStop}%
\bibitem [{\citenamefont {Helbig}\ \emph {et~al.}(2020)\citenamefont {Helbig},
  \citenamefont {Hofmann}, \citenamefont {Imhof}, \citenamefont {Abdelghany},
  \citenamefont {Kiessling}, \citenamefont {Molenkamp}, \citenamefont {Lee},
  \citenamefont {Szameit}, \citenamefont {Greiter},\ and\ \citenamefont
  {Thomale}}]{helbig2020generalized}%
  \BibitemOpen
  \bibfield  {author} {\bibinfo {author} {\bibfnamefont {T.}~\bibnamefont
  {Helbig}}, \bibinfo {author} {\bibfnamefont {T.}~\bibnamefont {Hofmann}},
  \bibinfo {author} {\bibfnamefont {S.}~\bibnamefont {Imhof}}, \bibinfo
  {author} {\bibfnamefont {M.}~\bibnamefont {Abdelghany}}, \bibinfo {author}
  {\bibfnamefont {T.}~\bibnamefont {Kiessling}}, \bibinfo {author}
  {\bibfnamefont {L.}~\bibnamefont {Molenkamp}}, \bibinfo {author}
  {\bibfnamefont {C.}~\bibnamefont {Lee}}, \bibinfo {author} {\bibfnamefont
  {A.}~\bibnamefont {Szameit}}, \bibinfo {author} {\bibfnamefont
  {M.}~\bibnamefont {Greiter}},\ and\ \bibinfo {author} {\bibfnamefont
  {R.}~\bibnamefont {Thomale}},\ }\bibfield  {title} {\bibinfo {title}
  {Generalized bulk-boundary correspondence in non-hermitian topolectrical
  circuits},\ }\href@noop {} {\bibfield  {journal} {\bibinfo  {journal} {Nature
  Physics}\ }\textbf {\bibinfo {volume} {16}},\ \bibinfo {pages} {747–750}
  (\bibinfo {year} {2020})}\BibitemShut {NoStop}%
\bibitem [{\citenamefont {Xiao}\ \emph {et~al.}(2020)\citenamefont {Xiao},
  \citenamefont {Deng}, \citenamefont {Wang}, \citenamefont {Zhu},
  \citenamefont {Wang}, \citenamefont {Yi},\ and\ \citenamefont
  {Xue}}]{xiao2020non}%
  \BibitemOpen
  \bibfield  {author} {\bibinfo {author} {\bibfnamefont {L.}~\bibnamefont
  {Xiao}}, \bibinfo {author} {\bibfnamefont {T.}~\bibnamefont {Deng}}, \bibinfo
  {author} {\bibfnamefont {K.}~\bibnamefont {Wang}}, \bibinfo {author}
  {\bibfnamefont {G.}~\bibnamefont {Zhu}}, \bibinfo {author} {\bibfnamefont
  {Z.}~\bibnamefont {Wang}}, \bibinfo {author} {\bibfnamefont {W.}~\bibnamefont
  {Yi}},\ and\ \bibinfo {author} {\bibfnamefont {P.}~\bibnamefont {Xue}},\
  }\bibfield  {title} {\bibinfo {title} {Non-hermitian bulk–boundary
  correspondence in quantum dynamics},\ }\href@noop {} {\bibfield  {journal}
  {\bibinfo  {journal} {Nature Physics}\ }\textbf {\bibinfo {volume} {16}},\
  \bibinfo {pages} {761} (\bibinfo {year} {2020})}\BibitemShut {NoStop}%
\bibitem [{\citenamefont {Ghatak}\ \emph {et~al.}(2020)\citenamefont {Ghatak},
  \citenamefont {Brandenbourger}, \citenamefont {van Wezel},\ and\
  \citenamefont {Coulais}}]{ghatak2020observation}%
  \BibitemOpen
  \bibfield  {author} {\bibinfo {author} {\bibfnamefont {A.}~\bibnamefont
  {Ghatak}}, \bibinfo {author} {\bibfnamefont {M.}~\bibnamefont
  {Brandenbourger}}, \bibinfo {author} {\bibfnamefont {J.}~\bibnamefont {van
  Wezel}},\ and\ \bibinfo {author} {\bibfnamefont {C.}~\bibnamefont
  {Coulais}},\ }\bibfield  {title} {\bibinfo {title} {Observation of
  non-hermitian topology and its bulk--edge correspondence in an active
  mechanical metamaterial},\ }\href@noop {} {\bibfield  {journal} {\bibinfo
  {journal} {Proceedings of the National Academy of Sciences}\ }\textbf
  {\bibinfo {volume} {117}},\ \bibinfo {pages} {29561} (\bibinfo {year}
  {2020})}\BibitemShut {NoStop}%
\bibitem [{\citenamefont {Wang}\ \emph {et~al.}(2021)\citenamefont {Wang},
  \citenamefont {Dutt}, \citenamefont {Yang}, \citenamefont {Wojcik},
  \citenamefont {Vu{\v{c}}kovi{\'c}},\ and\ \citenamefont
  {Fan}}]{wang2021generating}%
  \BibitemOpen
  \bibfield  {author} {\bibinfo {author} {\bibfnamefont {K.}~\bibnamefont
  {Wang}}, \bibinfo {author} {\bibfnamefont {A.}~\bibnamefont {Dutt}}, \bibinfo
  {author} {\bibfnamefont {K.~Y.}\ \bibnamefont {Yang}}, \bibinfo {author}
  {\bibfnamefont {C.~C.}\ \bibnamefont {Wojcik}}, \bibinfo {author}
  {\bibfnamefont {J.}~\bibnamefont {Vu{\v{c}}kovi{\'c}}},\ and\ \bibinfo
  {author} {\bibfnamefont {S.}~\bibnamefont {Fan}},\ }\bibfield  {title}
  {\bibinfo {title} {Generating arbitrary topological windings of a
  non-hermitian band},\ }\href@noop {} {\bibfield  {journal} {\bibinfo
  {journal} {Science}\ }\textbf {\bibinfo {volume} {371}},\ \bibinfo {pages}
  {1240} (\bibinfo {year} {2021})}\BibitemShut {NoStop}%
\bibitem [{\citenamefont {Li}\ \emph {et~al.}(2021{\natexlab{a}})\citenamefont
  {Li}, \citenamefont {Mu}, \citenamefont {Lee},\ and\ \citenamefont
  {Gong}}]{Li2021}%
  \BibitemOpen
  \bibfield  {author} {\bibinfo {author} {\bibfnamefont {L.}~\bibnamefont
  {Li}}, \bibinfo {author} {\bibfnamefont {S.}~\bibnamefont {Mu}}, \bibinfo
  {author} {\bibfnamefont {C.~H.}\ \bibnamefont {Lee}},\ and\ \bibinfo {author}
  {\bibfnamefont {J.}~\bibnamefont {Gong}},\ }\bibfield  {title} {\bibinfo
  {title} {Quantized classical response from spectral winding topology},\
  }\href@noop {} {\bibfield  {journal} {\bibinfo  {journal} {Nature
  communications}\ }\textbf {\bibinfo {volume} {12}},\ \bibinfo {pages} {5294}
  (\bibinfo {year} {2021}{\natexlab{a}})}\BibitemShut {NoStop}%
\bibitem [{\citenamefont {Budich}\ and\ \citenamefont
  {Bergholtz}(2020)}]{budich2020sensor}%
  \BibitemOpen
  \bibfield  {author} {\bibinfo {author} {\bibfnamefont {J.~C.}\ \bibnamefont
  {Budich}}\ and\ \bibinfo {author} {\bibfnamefont {E.~J.}\ \bibnamefont
  {Bergholtz}},\ }\bibfield  {title} {\bibinfo {title} {Non-hermitian
  topological sensors},\ }\href
  {https://doi.org/10.1103/PhysRevLett.125.180403} {\bibfield  {journal}
  {\bibinfo  {journal} {Phys. Rev. Lett.}\ }\textbf {\bibinfo {volume} {125}},\
  \bibinfo {pages} {180403} (\bibinfo {year} {2020})}\BibitemShut {NoStop}%
\bibitem [{\citenamefont {McDonald}\ and\ \citenamefont
  {Clerk}(2020)}]{mcdonald2020exponentially}%
  \BibitemOpen
  \bibfield  {author} {\bibinfo {author} {\bibfnamefont {A.}~\bibnamefont
  {McDonald}}\ and\ \bibinfo {author} {\bibfnamefont {A.~A.}\ \bibnamefont
  {Clerk}},\ }\bibfield  {title} {\bibinfo {title} {Exponentially-enhanced
  quantum sensing with non-hermitian lattice dynamics},\ }\href@noop {}
  {\bibfield  {journal} {\bibinfo  {journal} {Nature communications}\ }\textbf
  {\bibinfo {volume} {11}},\ \bibinfo {pages} {5382} (\bibinfo {year}
  {2020})}\BibitemShut {NoStop}%
\bibitem [{\citenamefont {Li}\ \emph {et~al.}(2021{\natexlab{b}})\citenamefont
  {Li}, \citenamefont {Lee},\ and\ \citenamefont {Gong}}]{li2021impurity}%
  \BibitemOpen
  \bibfield  {author} {\bibinfo {author} {\bibfnamefont {L.}~\bibnamefont
  {Li}}, \bibinfo {author} {\bibfnamefont {C.~H.}\ \bibnamefont {Lee}},\ and\
  \bibinfo {author} {\bibfnamefont {J.}~\bibnamefont {Gong}},\ }\bibfield
  {title} {\bibinfo {title} {Impurity induced scale-free localization},\
  }\href@noop {} {\bibfield  {journal} {\bibinfo  {journal} {Communications
  Physics}\ }\textbf {\bibinfo {volume} {4}},\ \bibinfo {pages} {1} (\bibinfo
  {year} {2021}{\natexlab{b}})}\BibitemShut {NoStop}%
\bibitem [{\citenamefont {Guo}\ \emph {et~al.}(2021)\citenamefont {Guo},
  \citenamefont {Liu}, \citenamefont {Zhao}, \citenamefont {Liu},\ and\
  \citenamefont {Chen}}]{guo2021exact}%
  \BibitemOpen
  \bibfield  {author} {\bibinfo {author} {\bibfnamefont {C.-X.}\ \bibnamefont
  {Guo}}, \bibinfo {author} {\bibfnamefont {C.-H.}\ \bibnamefont {Liu}},
  \bibinfo {author} {\bibfnamefont {X.-M.}\ \bibnamefont {Zhao}}, \bibinfo
  {author} {\bibfnamefont {Y.}~\bibnamefont {Liu}},\ and\ \bibinfo {author}
  {\bibfnamefont {S.}~\bibnamefont {Chen}},\ }\bibfield  {title} {\bibinfo
  {title} {Exact solution of non-hermitian systems with generalized boundary
  conditions: Size-dependent boundary effect and fragility of the skin
  effect},\ }\href {https://doi.org/10.1103/PhysRevLett.127.116801} {\bibfield
  {journal} {\bibinfo  {journal} {Phys. Rev. Lett.}\ }\textbf {\bibinfo
  {volume} {127}},\ \bibinfo {pages} {116801} (\bibinfo {year}
  {2021})}\BibitemShut {NoStop}%
\bibitem [{\citenamefont {Li}\ \emph {et~al.}(2020{\natexlab{b}})\citenamefont
  {Li}, \citenamefont {Lee}, \citenamefont {Mu},\ and\ \citenamefont
  {Gong}}]{li2020critical}%
  \BibitemOpen
  \bibfield  {author} {\bibinfo {author} {\bibfnamefont {L.}~\bibnamefont
  {Li}}, \bibinfo {author} {\bibfnamefont {C.~H.}\ \bibnamefont {Lee}},
  \bibinfo {author} {\bibfnamefont {S.}~\bibnamefont {Mu}},\ and\ \bibinfo
  {author} {\bibfnamefont {J.}~\bibnamefont {Gong}},\ }\bibfield  {title}
  {\bibinfo {title} {Critical non-hermitian skin effect},\ }\href@noop {}
  {\bibfield  {journal} {\bibinfo  {journal} {Nature communications}\ }\textbf
  {\bibinfo {volume} {11}} (\bibinfo {year} {2020}{\natexlab{b}})}\BibitemShut
  {NoStop}%
\bibitem [{\citenamefont {Liu}\ \emph {et~al.}(2020)\citenamefont {Liu},
  \citenamefont {Zhang}, \citenamefont {Yang},\ and\ \citenamefont
  {Chen}}]{liu2020helical}%
  \BibitemOpen
  \bibfield  {author} {\bibinfo {author} {\bibfnamefont {C.-H.}\ \bibnamefont
  {Liu}}, \bibinfo {author} {\bibfnamefont {K.}~\bibnamefont {Zhang}}, \bibinfo
  {author} {\bibfnamefont {Z.}~\bibnamefont {Yang}},\ and\ \bibinfo {author}
  {\bibfnamefont {S.}~\bibnamefont {Chen}},\ }\bibfield  {title} {\bibinfo
  {title} {Helical damping and anomalous critical non-hermitian skin effect},\
  }\href@noop {} {\bibfield  {journal} {\bibinfo  {journal} {arXiv preprint
  arXiv:2005.02617}\ } (\bibinfo {year} {2020})}\BibitemShut {NoStop}%
\bibitem [{\citenamefont {Mu}\ \emph {et~al.}()\citenamefont {Mu},
  \citenamefont {Zhou}, \citenamefont {Li},\ and\ \citenamefont
  {Gong}}]{mu2021nonhermitian}%
  \BibitemOpen
  \bibfield  {author} {\bibinfo {author} {\bibfnamefont {S.}~\bibnamefont
  {Mu}}, \bibinfo {author} {\bibfnamefont {L.}~\bibnamefont {Zhou}}, \bibinfo
  {author} {\bibfnamefont {L.}~\bibnamefont {Li}},\ and\ \bibinfo {author}
  {\bibfnamefont {J.}~\bibnamefont {Gong}},\ }\bibfield  {title} {\bibinfo
  {title} {Non-hermitian pseudo mobility edge in a coupled chain system},\
  }\href@noop {} {\ }\Eprint {https://arxiv.org/abs/2111.11914v1}
  {2111.11914v1} \BibitemShut {NoStop}%
\bibitem [{\citenamefont {Wiersig}(2014)}]{jan2014enhancing}%
  \BibitemOpen
  \bibfield  {author} {\bibinfo {author} {\bibfnamefont {J.}~\bibnamefont
  {Wiersig}},\ }\bibfield  {title} {\bibinfo {title} {Enhancing the sensitivity
  of frequency and energy splitting detection by using exceptional points:
  Application to microcavity sensors for single-particle detection},\ }\href
  {https://doi.org/10.1103/PhysRevLett.112.203901} {\bibfield  {journal}
  {\bibinfo  {journal} {Phys. Rev. Lett.}\ }\textbf {\bibinfo {volume} {112}},\
  \bibinfo {pages} {203901} (\bibinfo {year} {2014})}\BibitemShut {NoStop}%
\bibitem [{\citenamefont {Hodaei}\ \emph {et~al.}(2017)\citenamefont {Hodaei},
  \citenamefont {Hassan}, \citenamefont {Wittek}, \citenamefont
  {Garcia-Gracia}, \citenamefont {El-Ganainy}, \citenamefont
  {Christodoulides},\ and\ \citenamefont {Khajavikhan}}]{hodaei2017enhanced}%
  \BibitemOpen
  \bibfield  {author} {\bibinfo {author} {\bibfnamefont {H.}~\bibnamefont
  {Hodaei}}, \bibinfo {author} {\bibfnamefont {A.~U.}\ \bibnamefont {Hassan}},
  \bibinfo {author} {\bibfnamefont {S.}~\bibnamefont {Wittek}}, \bibinfo
  {author} {\bibfnamefont {H.}~\bibnamefont {Garcia-Gracia}}, \bibinfo {author}
  {\bibfnamefont {R.}~\bibnamefont {El-Ganainy}}, \bibinfo {author}
  {\bibfnamefont {D.~N.}\ \bibnamefont {Christodoulides}},\ and\ \bibinfo
  {author} {\bibfnamefont {M.}~\bibnamefont {Khajavikhan}},\ }\bibfield
  {title} {\bibinfo {title} {Enhanced sensitivity at higher-order exceptional
  points},\ }\href@noop {} {\bibfield  {journal} {\bibinfo  {journal} {Nature}\
  }\textbf {\bibinfo {volume} {548}},\ \bibinfo {pages} {187} (\bibinfo {year}
  {2017})}\BibitemShut {NoStop}%
\bibitem [{\citenamefont {Chen}\ \emph {et~al.}(2017)\citenamefont {Chen},
  \citenamefont {{\"O}zdemir}, \citenamefont {Zhao}, \citenamefont {Wiersig},\
  and\ \citenamefont {Yang}}]{chen2017exceptional}%
  \BibitemOpen
  \bibfield  {author} {\bibinfo {author} {\bibfnamefont {W.}~\bibnamefont
  {Chen}}, \bibinfo {author} {\bibfnamefont {{\c{S}}.~K.}\ \bibnamefont
  {{\"O}zdemir}}, \bibinfo {author} {\bibfnamefont {G.}~\bibnamefont {Zhao}},
  \bibinfo {author} {\bibfnamefont {J.}~\bibnamefont {Wiersig}},\ and\ \bibinfo
  {author} {\bibfnamefont {L.}~\bibnamefont {Yang}},\ }\bibfield  {title}
  {\bibinfo {title} {Exceptional points enhance sensing in an optical
  microcavity},\ }\href@noop {} {\bibfield  {journal} {\bibinfo  {journal}
  {Nature}\ }\textbf {\bibinfo {volume} {548}},\ \bibinfo {pages} {192}
  (\bibinfo {year} {2017})}\BibitemShut {NoStop}%
\bibitem [{\citenamefont {McDonald}\ \emph {et~al.}(2018)\citenamefont
  {McDonald}, \citenamefont {Pereg-Barnea},\ and\ \citenamefont
  {Clerk}}]{McDonald2018}%
  \BibitemOpen
  \bibfield  {author} {\bibinfo {author} {\bibfnamefont {A.}~\bibnamefont
  {McDonald}}, \bibinfo {author} {\bibfnamefont {T.}~\bibnamefont
  {Pereg-Barnea}},\ and\ \bibinfo {author} {\bibfnamefont {A.~A.}\ \bibnamefont
  {Clerk}},\ }\bibfield  {title} {\bibinfo {title} {Phase-dependent chiral
  transport and effective non-hermitian dynamics in a bosonic kitaev-majorana
  chain},\ }\href {https://doi.org/10.1103/PhysRevX.8.041031} {\bibfield
  {journal} {\bibinfo  {journal} {Phys. Rev. X}\ }\textbf {\bibinfo {volume}
  {8}},\ \bibinfo {pages} {041031} (\bibinfo {year} {2018})}\BibitemShut
  {NoStop}%
\bibitem [{\citenamefont {Wanjura}\ \emph {et~al.}(2020)\citenamefont
  {Wanjura}, \citenamefont {Brunelli},\ and\ \citenamefont
  {Nunnenkamp}}]{Wanjura2020}%
  \BibitemOpen
  \bibfield  {author} {\bibinfo {author} {\bibfnamefont {C.~C.}\ \bibnamefont
  {Wanjura}}, \bibinfo {author} {\bibfnamefont {M.}~\bibnamefont {Brunelli}},\
  and\ \bibinfo {author} {\bibfnamefont {A.}~\bibnamefont {Nunnenkamp}},\
  }\bibfield  {title} {\bibinfo {title} {Topological framework for directional
  amplification in driven-dissipative cavity arrays},\ }\href@noop {}
  {\bibfield  {journal} {\bibinfo  {journal} {Nature communications}\ }\textbf
  {\bibinfo {volume} {11}},\ \bibinfo {pages} {1} (\bibinfo {year}
  {2020})}\BibitemShut {NoStop}%
\bibitem [{\citenamefont {Xue}\ \emph {et~al.}(2020)\citenamefont {Xue},
  \citenamefont {Li}, \citenamefont {Hu}, \citenamefont {Song},\ and\
  \citenamefont {Wang}}]{xue2020non}%
  \BibitemOpen
  \bibfield  {author} {\bibinfo {author} {\bibfnamefont {W.-T.}\ \bibnamefont
  {Xue}}, \bibinfo {author} {\bibfnamefont {M.-R.}\ \bibnamefont {Li}},
  \bibinfo {author} {\bibfnamefont {Y.-M.}\ \bibnamefont {Hu}}, \bibinfo
  {author} {\bibfnamefont {F.}~\bibnamefont {Song}},\ and\ \bibinfo {author}
  {\bibfnamefont {Z.}~\bibnamefont {Wang}},\ }\bibfield  {title} {\bibinfo
  {title} {Non-hermitian band theory of directional amplification},\
  }\href@noop {} {\bibfield  {journal} {\bibinfo  {journal} {arXiv preprint
  arXiv:2004.09529}\ } (\bibinfo {year} {2020})}\BibitemShut {NoStop}%
\bibitem [{Note1()}]{Note1}%
  \BibitemOpen
  \bibinfo {note} {For $|\nu _\alpha |>1$, the single element shall be replaced
  by the determinant of its off-diagonal block when calculating the responses
  in Eq. \protect \textup {\hbox {\mathsurround \z@ \protect \normalfont
  (\ignorespaces \ref {eq:response}\unskip \@@italiccorr )}} \cite
  {Li2021}}\BibitemShut {NoStop}%
\bibitem [{sup()}]{suppmat}%
  \BibitemOpen
  \bibfield  {title} {\bibinfo {title} {Supplemental materials},\ }\href@noop
  {} {\bibinfo  {journal} {Supplemental Materials}\ }\BibitemShut {NoStop}%
\bibitem [{\citenamefont {Koch}\ and\ \citenamefont
  {Budich}(2020)}]{koch2020bulk}%
  \BibitemOpen
\bibfield  {journal} {  }\bibfield  {author} {\bibinfo {author} {\bibfnamefont
  {R.}~\bibnamefont {Koch}}\ and\ \bibinfo {author} {\bibfnamefont {J.~C.}\
  \bibnamefont {Budich}},\ }\bibfield  {title} {\bibinfo {title} {Bulk-boundary
  correspondence in non-hermitian systems: stability analysis for generalized
  boundary conditions},\ }\href@noop {} {\bibfield  {journal} {\bibinfo
  {journal} {The European Physical Journal D}\ }\textbf {\bibinfo {volume}
  {74}},\ \bibinfo {pages} {1} (\bibinfo {year} {2020})}\BibitemShut {NoStop}%
\bibitem [{\citenamefont {Kunst}\ \emph {et~al.}(2018)\citenamefont {Kunst},
  \citenamefont {Edvardsson}, \citenamefont {Budich},\ and\ \citenamefont
  {Bergholtz}}]{kunst2018biorthogonal}%
  \BibitemOpen
  \bibfield  {author} {\bibinfo {author} {\bibfnamefont {F.~K.}\ \bibnamefont
  {Kunst}}, \bibinfo {author} {\bibfnamefont {E.}~\bibnamefont {Edvardsson}},
  \bibinfo {author} {\bibfnamefont {J.~C.}\ \bibnamefont {Budich}},\ and\
  \bibinfo {author} {\bibfnamefont {E.~J.}\ \bibnamefont {Bergholtz}},\
  }\bibfield  {title} {\bibinfo {title} {Biorthogonal bulk-boundary
  correspondence in non-hermitian systems},\ }\href
  {https://doi.org/10.1103/PhysRevLett.121.026808} {\bibfield  {journal}
  {\bibinfo  {journal} {Phys. Rev. Lett.}\ }\textbf {\bibinfo {volume} {121}},\
  \bibinfo {pages} {026808} (\bibinfo {year} {2018})}\BibitemShut {NoStop}%
\bibitem [{\citenamefont {Chang}\ \emph {et~al.}(2020)\citenamefont {Chang},
  \citenamefont {You}, \citenamefont {Wen},\ and\ \citenamefont
  {Ryu}}]{chang2020entanglement}%
  \BibitemOpen
  \bibfield  {author} {\bibinfo {author} {\bibfnamefont {P.-Y.}\ \bibnamefont
  {Chang}}, \bibinfo {author} {\bibfnamefont {J.-S.}\ \bibnamefont {You}},
  \bibinfo {author} {\bibfnamefont {X.}~\bibnamefont {Wen}},\ and\ \bibinfo
  {author} {\bibfnamefont {S.}~\bibnamefont {Ryu}},\ }\bibfield  {title}
  {\bibinfo {title} {Entanglement spectrum and entropy in topological
  non-hermitian systems and nonunitary conformal field theory},\ }\href@noop {}
  {\bibfield  {journal} {\bibinfo  {journal} {Physical Review Research}\
  }\textbf {\bibinfo {volume} {2}},\ \bibinfo {pages} {033069} (\bibinfo {year}
  {2020})}\BibitemShut {NoStop}%
\bibitem [{\citenamefont {Li}\ and\ \citenamefont {Lee}()}]{li2021non}%
  \BibitemOpen
  \bibfield  {author} {\bibinfo {author} {\bibfnamefont {L.}~\bibnamefont
  {Li}}\ and\ \bibinfo {author} {\bibfnamefont {C.~H.}\ \bibnamefont {Lee}},\
  }\bibfield  {title} {\bibinfo {title} {Non-hermitian pseudo-gaps},\
  }\href@noop {} {\ }\Eprint {https://arxiv.org/abs/2106.02995v1}
  {2106.02995v1} \BibitemShut {NoStop}%
\end{thebibliography}
%

\end{document}